\documentstyle[preprint,aps]{revtex}
\input epsf
\tightenlines
\begin{document}

\draft

\title{The general cancellation of ladder graphs at finite temperature}

\author{M.~E.~Carrington $^a$ and R.~Kobes $^b$}

\address{
  $^a$ Department of Physics \\
  Winnipeg Institute of Theoretical Physics\\
  University of Brandon\\ Brandon, Manitoba, R7A 6A9\\ Canada\\
  $^b$ Department of Physics \\
  Winnipeg Institute of Theoretical Physics\\
  University of Winnipeg\\ Winnipeg, Manitoba, R3B 2E9\\ Canada}

\date{\today}

\maketitle

\begin{abstract}
  In some cases, an important example being at finite temperature,
  extreme infrared, collinear, or light--cone behaviour may
  cause the usual loop expansion to break down. 
  For some of these cases higher order ladder graphs can become
  important. In an earlier paper it was shown that, given a particular
  relation between a vertex and a self--energy
  function, the resummation of the ladder graphs simplifies significantly
  when other types of graphs are included in a consistent effective
  expansion. In this paper we show that this assumed relation is valid
  for a large class of vertex and self--energy functions at finite
  temperature.
\end{abstract}
\vspace{3cm}
\pacs{PACS numbers: 11.10Wx, 11.15Tk, 11.55Fv}

\narrowtext

 %%%%%%%%%%%%%%%%%%%%%%%%%%%%%%%%%%%%%%%%%%%%%%%%%%%%%%%%%%%%%%%%%%
\section{Introduction}
\label{sec1}
%%%%%%%%%%%%%%%%%%%%%%%%%%%%%%%%%%%%%%%%%%%%%%%%%%%%%%%%%%%%%%%%%%
Particularly at finite temperature, infrared divergences can cause
the loop expansion to fail. In many cases this problem can be resolved
by using an effective expansion based on the hard thermal loops
\cite{r8,klim,wel,brat,wong,tay}. This expansion does not solve all
such problems, however; we mention in this context the damping
rate of a fast fermion, where a self--consistent
calculational scheme outside of the hard thermal loop expansion
has been used \cite{smilga,piz},
processes sensitive to the behaviour near the light--cone, where
an ``improved'' hard thermal loop expansion has been proposed
\cite{flech,kraemmer}, the calculation of the
pressure \cite{pressure}, and the problem of the effective
potential in electro--weak theory \cite{ep}.
\par
In this paper we consider perturbation expansions beyond the
loop expansion which include
ladder graphs. These graphs, which are not included in the
hard thermal loop expansion, become important in certain cases
sensitive to the infrared and/or light--cone limits
\cite{smilga,kraemmer,pressure,meg}, and also arise in the
context of the eikonal expansion of gauge theories \cite{cornwall,hou}.
We show for the calculation of a self--energy that there is an
effective expansion for these graphs
which arises as the iterative solution of a type of
Schwinger--Dyson equation. However, in general other types of graphs
must be included in a consistent effective expansion.
In this regard, when a relation of the generic form
\begin{equation}
  \Gamma(K, P) \approx -ig\frac{1}{K^2+2K\cdot P}
  \left[\Pi(P) -\Pi(K+P)\right]
\end{equation}
between the vertex function $\Gamma(K, P)$ and a self--energy function
$\Pi(P)$ holds, the inclusion of the non--ladder graphs
gives rise to a significant simplification in the expression for the
self--energy \cite{us}. We show here that this relation is valid
for a large class of vertex and self--energy functions at finite
temperature. Finally, if certain assumptions are made on the
relative size of the self--energy function $\Pi(P)$, we show
that the ladder graphs actually cancel against other types of terms.
  
%%%%%%%%%%%%%%%%%%%%%%%%%%%%%%%%%%%%%%%%%%%%%%%%%%%%%%%%%%%%%%%%%
\section{Ladder Graphs}
\label{sec2}
%%%%%%%%%%%%%%%%%%%%%%%%%%%%%%%%%%%%%%%%%%%%%%%%%%%%%%%%%%%%%%%%%%
We begin by showing under what circumstances ladder graphs are
important.
We work here with a scalar $\phi^3$ theory at zero temperature,
but the results
generalize straightforwardly for other theories and to finite
temperature.
Consider first the graph of Fig.\ref{ladderfig}, which is given by
\begin{eqnarray}
  \label{ladder}
  -i\Pi(K)& =& (-ig)^6\int\, dR_1\, dR_2\, dP \,
  D(P+R_1+R_2)D(P+R_1+R_2+K)D(R_1)
  D(P+R_2)\nonumber\\& &D(P+R_2+K)D(R_2)
  D(P)D(P+K),
\end{eqnarray}
where $D(K) = 1 / (K^2+i\epsilon)$
and $K=(k_0, {\vec k})$. In cases where the loop
expansion is valid this graph would be suppressed by a factor of $g^4$
relative to the one--loop graph of Fig.~\ref{oneloop}. However, especially
at finite temperature, circumstances may arise where this is not the case.
Let us split two of the propagators in Eq.~[\ref{ladder}] as
\begin{equation}
  D(P+R_2)D(P+R_2+K) = 
  \frac{D(P+R_2)-D(P+R_2+K)}{K^2+2K\cdot (P+R_2)},
\end{equation}
and furthermore consider the infrared limit
$2K\cdot R_2 \ll (K^2 + 2K\cdot P)$, whereby
this splitting is approximated by
\begin{equation}
  \label{split}
  D(P+R_2)D(P+R_2+K) \approx 
  \frac{D(P+R_2)-D(P+R_2+K)}{K^2+2K\cdot P},
\end{equation}
We perform an analogous split for $D(P+R_1+R_2)D(P+R_1+R_2+K)$.
Such approximations in Eq.~[\ref{ladder}] lead to
\begin{eqnarray}
  \label{splitladder}
  -i\Pi(K) &\approx& (-ig)^6\int\, dR_1\, dR_2\, dP\,
  \frac{D(P+R_1+R_2)-D(P+R_1+R_2+K)}{K^2+2K\cdot P}
  D(R_1)\nonumber\\
  & &\frac{D(P+R_2)-D(P+R_2+K)}{K^2+2K\cdot P}
  D(R_2)D(P)D(P+K).
\end{eqnarray}
Now, if it happens that a
region of phase space exists where $(K^2+2K\cdot P)$ is sufficiently
small (for example, at finite temperature when
$(K^2+2K\cdot P) \sim O(g^2T^2)$), then
a factor of $g^4$ arises in the denominator of Eq.~[\ref{splitladder}] which
would cancel a factor of $g^4$ in the numerator. This would lead  to a
situation where the ladder graph of Fig.~\ref{ladderfig} is of the
same order as the one--loop term of Fig.~\ref{oneloop}, signaling the
breakdown of the loop expansion.
\par
The breakdown of the loop expansion in this manner is due to the importance
of the ladder graph in Fig.~\ref{ladderfig} and similar higher loop ladder
graphs.
Higher loop ``crossed ladders'' such as that illustrated in
Fig.~\ref{unladderfig}, and given by
\begin{eqnarray}
  -i\Pi(K)& =& (-ig)^6\int\, dR_1\, dR_2\, dP\,
  D(P+R_1+R_2)D(P+R_1+R_2+K)D(R_1)
  D(P+R_2)\nonumber\\ & &D(P+R_1+K)D(R_2)
  D(P)D(P+K),
\end{eqnarray}
do not contribute in the same way as the ladder graphs. This can
be seen as follows.
The product of propagators $D(P+R_1+R_2)D(P+R_1+R_2+K)$ can be split
along the lines of Eq.~[\ref{split}], the product
$D(P+R_2)D(P+R_1+K)$ would be split as
\begin{equation}
  D(P+R_2)D(P+R_1+K) =
  \frac{D(P+R_2)-D(P+R_1+K)}
  {K^2+2K\cdot (P+R_1) + (P+R_1)^2 - (P+R_2)^2}.
\end{equation}
The infrared limit $2K\cdot R_1 \ll (K^2 + 2K\cdot P)$ then would produce
\begin{equation}
  D(P+R_2)D(P+R_1+K) \approx
  \frac{D(P+R_2)-D(P+R_1+K)}
  {K^2+2K\cdot P + (P+R_1)^2 - (P+R_2)^2},
\end{equation}
which, in the region $(K^2+2K\cdot P) \sim O(g^2T^2)$, would not
by itself lead to a
cancellation of a factor of $g^2$ in the numerator due to the presence of the
$(P+R_1)^2 - (P+R_2)^2$ term. One could try to get such
a cancellation by furthermore restricting the phase space so that $P\cdot R_i$
and $R_i^2$ ($i=1,2$) are sufficiently small, but this introduces extra factors
of $g$ in the numerator coming from the momentum integral over $P$.
The conclusion one draws is that in this infrared limit such crossed
graphs are suppressed relative to the ladder graphs.
%%%%%%%%%%%%%%%%%%%%%%%%%%%%%%%%%%%%%%%%%%%%%%%%%%%%%%%%%%%%%%%%%%
\section{Ladder resummation}
\label{sec3}
%%%%%%%%%%%%%%%%%%%%%%%%%%%%%%%%%%%%%%%%%%%%%%%%%%%%%%%%%%%%%%%%%%
In this section we describe a method for including the ladder graphs
discussed in the previous section in an effective expansion. To this
end, we first consider the one--loop vertex of Fig.~\ref{vertexfig}.
The expression for this graph is
\begin{equation}
  \Gamma(K, P) = (-ig)^3 \int\, dR\, D(R)D(R+P)D(K+P+R).
\end{equation}
We split the two propagators $D(R+P)D(K+P+R)$ as in Eq.~[\ref{split}]
and
use the approximation $2K\cdot R \ll (K^2 + 2K\cdot P)$. Comparing
the result of this operation, 
\begin{equation}
  \Gamma(K, P) \approx (-ig)^3 \frac{1}{K^2+2K\cdot P}
  \int dR \,D(R)\left[D(R+P)-D(K+P+R)\right],
\end{equation}
to the one--loop self--energy graph of Fig.~\ref{oneloop},
\begin{equation}
  \label{oneloopselfenergy}
  -i\Pi(P) = -(-ig)^2 \int\, dR\, D(R)D(R+P),
\end{equation}
we find the relation
\begin{equation}
  \label{relation}
  \Gamma(K, P) \approx -ig\frac{1}{K^2+2K\cdot P}
  \left[\Pi(P) -\Pi(K+P)\right].
\end{equation}
Note that, due to the absence of an $i\epsilon$ in the
denominator, we must assume in
this expression that we  are in a region of phase space where $K^2+2K\cdot P$
does not vanish.
\par
Similar results as in Eq.~[\ref{relation}] hold in gauge theories. For
example, in fermionic $QED$, the analogous result is
\cite{meg}
\begin{equation}
  \label{sqedrelation}
  \Gamma_\mu(K, P) \approx -ig\frac{K_\mu + 2P_\mu}{K^2+2K\cdot P}
  \left[\Pi(P) -\Pi(K+P)\right],
\end{equation}
where the one--loop self--energy graph is shown in Fig.~\ref{oneloop}
(with the photon line having momentum $R$).
This relation illustrates the connection between this approximation for
the vertex function and gauge invariance: contracting
Eq.~[\ref{sqedrelation}] with $K^\mu$ leads to
\begin{equation}
  K\cdot \Gamma(K, P) = -ig\left[\Pi(P) -\Pi(K+P)\right],
\end{equation}
which of course is the Ward identity for the vertex function.
In a sense, this approximation for the vertex function is
equivalent to ``solving'' the Ward identity for this function
in the infrared limit.
\par
These considerations allow us to construct a method for
generating the ladder graphs. To this end, consider the
equation indicated in
Fig.~\ref{semiselfenergy}  for the
full self--energy.: note that, due to the use of the
bare rather than full propagators on the internal lines, 
this is a ``partial'' Schwinger--Dyson equation. In momentum
space this equation is
\begin{equation}
  \label{semisdequation}
  -i\Pi(K) = ig\int\, dR\,\Gamma(K, R)
  D(R)D(K+R),
\end{equation}
where $\Gamma(K,R)$ is the full three--point vertex.
Use of Eq.~(\ref{relation}) for the full vertex function
in terms of the self--energy $\Pi$ then generates a
 perturbative summation of ladder
graphs, such as in Fig.~\ref{ladderfig}, under the appropriate
approximations of small loop momenta used in the derivation.
To illustrate this point, if for the first iteration we use $\Gamma(K,R)$ of
Eq.~(\ref{relation}) with the one--loop self--energy of
Fig.~\ref{oneloop} given in Eq.~(\ref{oneloopselfenergy}),
we find
\begin{eqnarray}
  \label{graph1}
   -i\Pi(K)& =& (-ig)^2\int\, dR\,
   \frac{\left[\Pi(R) -\Pi(K+R)\right]}{K^2+2K\cdot R}
   D(R)D(K+R)\nonumber\\
   &=& (-ig)^2\int\, dR\,\frac{ D(R)D(K+R)}{K^2+2K\cdot R}
   i(-ig)^2\int\,dR^\prime\,\nonumber\\
   & &\qquad\left[
     D(R^\prime)D(R+R^\prime)-
     D(R^\prime)D(K+R+R^\prime)\right]\nonumber\\
   &=&i(-ig)^4\int\,dR\,dR^\prime\,
   D(R)D(K+R)D(R^\prime)D(R+R^\prime)
   D(K+R+R^\prime),
\end{eqnarray}
where we assumed $2K\cdot R^\prime \ll (K^2 + 2K\cdot R)$.
This expression corresponds to the ladder graph of
Fig.~\ref{firstladder}.
\par
We note that in this derivation there are two competing limits
for the size of $K^2 + 2K\cdot R$: one is the infrared limit
of $2K\cdot R^\prime \ll (K^2 + 2K\cdot R)$, while the other
is the fact that $K^2 + 2K\cdot R$ must be sufficiently small
so as to make the ladder graphs important (see
Eq.~(\ref{splitladder})). It is only in special cases, such
as some finite temperature examples, where both of these
criteria may be simultaneously satisfied.
\par
The particular resummation of ladder graphs indicated above is for
graphs with bare internal lines and vertices. The question
then arises as to how one generates such expansions when
some combination of the internal lines and vertices in the ladder
graphs are replaced by
their effective counterparts, as in  in Figs.~\ref{fig3}--\ref{fig10}.
 This question may be addressed in the same way as
for bare vertices and internal lines.
We first show that in the limit $K\rightarrow 0$ the vertices and 
polarization tensors in Figs.~\ref{fig3}--\ref{fig10} satisfy 
Eq.~[\ref{relation}].  Using this result, the iteration of the partial
Schwinger--Dyson equation~(\ref{semisdequation}) 
shown in Fig.~\ref{semiselfenergy}
can then be used to generate the effective resummation of ladder
graphs for the self--energy.
\par
For example, consider the vertex of
Fig.~\ref{fig3} with one corrected internal line:
\begin{equation}
  \Gamma(K, P) = -i(-ig)^3 \int\, dR\, G(R)D(R+P)D(K+P+R),
\end{equation}
where $ G(R) = 1 / (R^2-\Pi(R)+i\epsilon) $
is the full propagator. We split the two propagators
$D(R+P)D(K+P+R)$ as in Eq.~[\ref{split}] and
use the approximation $2K\cdot R \ll (K^2 + 2K\cdot P)$, whereby this equation
becomes
\begin{equation}
  \Gamma(K, P) \approx -i(-ig)^3 \frac{1}{K^2+2K\cdot P}
  \int dR\,\, G(R)\left[D(R+P)-D(K+P+R)\right].
\end{equation}
Comparing this to the self--energy graph of Fig.~\ref{fig4}:
\begin{equation}
  -i\Pi(P) = - (-ig)^2 \int\, dR\, G(R)D(R+P),
\end{equation}
we find the relation Eq.~(\ref{relation}) is satisfied.
Use of this relation in the iteration of the partial Schwinger--Dyson
equation of Eq.~(\ref{semisdequation}) shown in Fig.~\ref{semiselfenergy}
would lead then to a series of ladder graphs
with effective propagators on the internal ``vertical'' lines,
such as in Fig.~\ref{laddercorrected}. To see this,
if for the first iteration we use $\Gamma(K,R)$ of
Eq.~(\ref{relation}) with the self--energy of
Fig.~\ref{fig4},
we find
\begin{eqnarray}
   -i\Pi(K)& =& (ig)^2\int\, dR\,
   \frac{\left[\Pi(R) -\Pi(K+R)\right]}{K^2+2K\cdot R}
   D(R)D(K+R)\nonumber\\
   &=& -(ig)^2\int\, dR\,\frac{ D(R)D(K+R)}{K^2+2K\cdot R}
   i(-ig)^2\int\,dR^\prime\,\nonumber\\
   & &\qquad\left[
     G(R^\prime)D(R+R^\prime)-
     G(R^\prime)D(K+R+R^\prime)\right]\nonumber\\
   &=&i(-ig)^4\int\,dR\,dR^\prime\,
   D(R)D(K+R)G(R^\prime)D(R+R^\prime)
   D(K+R+R^\prime),
\end{eqnarray}
where we assumed $2K\cdot R^\prime \ll (K^2 + 2K\cdot R)$.
This corresponds to the ladder graph of Fig.~\ref{laddercorrected}.
\par
We thus see that the ladder graphs may be generated by the iteration
of the partial Schwinger--Dyson equation of Fig.~\ref{semiselfenergy}.
However, for completeness the full Schwinger--Dyson equation of 
Fig.~\ref{fullselfenergy} should be considered. In Ref.\cite{us} it
was shown that this equation,
\begin{equation}
  \label{complete}
  -i\Pi(K) = ig\int\, dR\,\left[-ig+\Gamma(K, R)\right]
  G(R)G(K+R),
\end{equation}
where 
$G(K) = i / (K^2-\Pi(K)+i\epsilon)$
is the full propagator, simplifies to
\begin{equation}
  \label{simplify}
  -i\Pi(K) = 2(-ig)^2\int\, dR\, \left[\frac{i}{K^2+2K\cdot R}\right]
  \left[\frac{i}{R^2-\Pi(R)+i\epsilon}\right]
\end{equation}
if Eq.~(\ref{relation}) hold. 
 This simplification results from a
partial cancellation of ladder graphs against non--ladder graphs.
To see this cancellation explicitly, start with Eq.~(\ref{complete}) 
and rewrite $\Gamma(K,R)$ in terms of $\Pi(K)$ using Eq.~(\ref{relation}).
Taking as an example the one--loop expression of Fig.~\ref{oneloop},
we find the first iteration of  Eq.~(\ref{complete}) gives
\begin{equation}
  -i\Pi^{(1)}(K) = 2(-ig)^2\int\, dR\, \left[\frac{i}{K^2+2K\cdot R}\right]
  \left[\frac{i}{R^2+i\epsilon}\right],
\end{equation}
which corresponds to the one--loop graph of Fig.~\ref{oneloop}, which
is given in Eq.~(\ref{simplify}). 
At the next order of iteration we have two terms: one from the
ladder graph of Fig.~\ref{firstladder}, given in Eq.~(\ref{graph1}), and
the other from the two--loop graph of Fig.~\ref{cancel}:
\begin{equation}
   2(-ig)^2\int\, dR\, dR^\prime
   D(R)D(K+R)D(K+R+R^\prime)D(R^\prime)D(K+R).
\end{equation}
When these are combined in the infrared limit one obtains 
\begin{equation}
  2(-ig)^2\int\, dR\, \left[\frac{i}{K^2+2K\cdot R}\right]
  \frac{i}{R^2+i\epsilon}\left[-i\Pi(R)\right]\frac{i}{R^2+i\epsilon},
\end{equation}
which is just the second term of the expansion  in
terms of $\Pi$ of Eq.~(\ref{simplify}).
Thus, ladder and non--ladder graphs combine to
give a relatively simple expression. Further simplification is
possible if we assume in Eq.~(\ref{simplify}) that the contribution
of $\Pi(R)$ to the pole in the integral is negligible, 
so that this expression becomes
\begin{equation}
  -i\Pi(K) = 2(-ig)^2\int\, dR\, \left[\frac{i}{K^2+2K\cdot R}\right]
  \left[\frac{i}{R^2+i\epsilon}\right].
\end{equation}
This corresponds, in the infrared limit, to the one--loop graph
of Fig.~\ref{oneloop}. In this case one can say that the ladder
graph of Fig.~\ref{firstladder} has canceled exactly against the
non--ladder graph of Fig.~\ref{cancel}; such cancellation in some
particular examples has been found in Refs.\cite{smilga,kraemmer,meg}.
\par
Thus, we see that the relation of Eq.~(\ref{relation})
between the vertex function $\Gamma$ and self--energy $\Pi$ is critical
in two respects.
Firstly, it gives an expansion of the ladder graphs in
terms of the iteration of the partial Schwinger--Dyson equation
of Fig.~\ref{semiselfenergy}. Secondly, it demonstrates
 how ladder graphs
combine with other types of terms in the full Schwinger--Dyson equation of 
Fig.~\ref{fullselfenergy} to give a relatively simple expression. 
The rest of the paper
addresses the question of the validity of this relation at finite
temperature for more complicated cases than the ones just considered.
 In the next three sections we express the
2, 3, and 4--point functions in a particular real--time formalism
of finite temperature field theory useful for our purposes. We then
show how in this formalism the relation of Eq.~(\ref{relation})
is satisfied
at finite temperature for a wide class of vertex and self--energy
diagrams with corrected vertices and internal lines.
We end with some brief conclusions.
%%%%%%%%%%%%%%%%%%%%%%%%%%%%%%%%%%%%%%%%%%%%%%%%%%%%%%%%%%%%%%%%%%%
%%%%%%%%%%%%%%%%%%%%%%%%%%%%%%%%%%%%%%%%%%%%%%%%%%%%%%%%%%%%%%%%%%
\section{Propagator}
\label{sec4}
%%%%%%%%%%%%%%%%%%%%%%%%%%%%%%%%%%%%%%%%%%%%%%%%%%%%%%%%%%%%%%%%%%

We first consider the
propagator. In real time, the propagator has $2^2=4$ components, since
each of the two fields can take values on either branch of the
contour. Thus, the propagator can be written as a $2 \times 2 $ matrix
of the form
 \begin{equation}
 \label{2}
   D = \left(  \matrix {D_{11} & D_{12} \cr
                        D_{21} & D_{22} \cr} \right) \, ,
 \end{equation}
where $D_{11}$ is the propagator for fields moving along $C_1$,
$D_{12}$ is the propagator for fields moving from $C_1$ to $C_2$, etc.
The four components are given by
 \begin{eqnarray}
 \label{eq: compD}
   D_{11}(x-y) &=& -i\langle T(\phi(x) \phi(y))\rangle \, , \nonumber\\
   D_{12}(x-y) &=& -i\langle \phi(y) \phi(x) \rangle \, , \nonumber\\
   D_{21}(x-y) &=& -i\langle \phi(x) \phi(y)\rangle \, , \nonumber\\
   D_{22}(x-y) &=& -i\langle\tilde{T}(\phi(x)\phi(y))\rangle \, ,
 \end{eqnarray}
where $T$ is the usual time ordering operator, and $\tilde{T}$ is the
anti-chronological time ordering operator. These four components
satisfy,
 \begin{equation}
 \label{3}
   D_{11} - D_{12} - D_{21} + D_{22} = 0
 \end{equation}
as a consequence of the identity $\theta(x) + \theta(-x) =1$.

It is more useful to write the propagator in terms of the three functions
 \begin{eqnarray} \label{3a}
   D_R &=& D_{11} - D_{12} \, , \nonumber\\
   D_A &=& D_{11} - D_{21} \, , \nonumber\\
   D_F &=& D_{11} + D_{22} \, .
 \end{eqnarray}
$D_R$ and $D_A$ are the usual retarded and advanced propagators,
satisfying
 \begin{equation}
 \label{4}
   D_R(x-y)-D_A(x-y) = -i\langle [\phi(x),\phi(y)] \rangle\, ,
 \end{equation}
and $D_F$ is the symmetric combination
 \begin{equation}
 \label{4a}
   D_F(x-y) = -i\langle \{\phi(x),\phi(y)\} \rangle\, ,
 \end{equation}
which satisfies in momentum space
\begin{equation}
D_F(P) = (1+2n(p_0))(D_R(P) - D_A(P)),
\label{eq: 4b}
\end{equation}
where
 \begin{equation}
   D_{R,A}(P) = \frac{1}{(p_0\pm i\epsilon)^2 - {\vec p}^{\,2}-m^2}.
 \end{equation}

Equations~(\ref{3}),~(\ref{3a}) are inverted by
 \begin{eqnarray}
 \label{7}
   D_{11} &=& \frac{1}{2} (D_F + D_A + D_R) \, , \nonumber\\
   D_{12} &=& \frac{1}{2} (D_F  +D_A - D_R) \, , \nonumber\\
   D_{21} &=& \frac{1}{2} (D_F  -D_A + D_R) \, , \nonumber\\
   D_{22} &=& \frac{1}{2} (D_F  -D_A - D_R) \, .
 \end{eqnarray}
These equations can be written in a more convenient notation as \cite{Chou}
 \begin{equation}
 \label{eq: decompD1}
   2\,D = D_R {1\choose 1}{1\choose -1}
        + D_A {1\choose -1}{1\choose 1}
        + D_F {1\choose 1}{1\choose 1}
 \end{equation}
where the outer product of the column vectors is to be taken.
\par
Similar relations can be obtained for the polarization tensor.  The
polarization tensor is the 1PI
two-point function and is obtained by amputating the external legs from
the propagator. The Dyson equation gives
 \begin{equation}
 \label{8}
   iD(p) = iD_0(p) + iD_0(p) \, \bigl(-i\Pi(p)\bigr)\, iD(p) \, .
 \end{equation}
The analogues of~(\ref{3}) and~(\ref{3a}) are
 \begin{eqnarray}
   \Pi_R &=& \Pi_{11} + \Pi_{12} \, , \nonumber\\
   \Pi_A &=& \Pi_{11} + \Pi_{21} \, , \nonumber\\
   \Pi_F &=& \Pi_{11} + \Pi_{22} \, ,
 \label{eq: physpi}
 \end{eqnarray}
and
 \begin{equation}
   \Pi_{11} + \Pi_{12} + \Pi_{21} + \Pi_{22} = 0 \, .
 \label{eq: circpi}
 \end{equation}
The analogues of~(\ref{eq: decompD1}) and~(\ref{eq: 4b}) are
 \begin{eqnarray}
 \label{eq: Pidecomp1}
   2\, \Pi (p) &=& \Pi_R(p) {1\choose -1} {1\choose 1}
             + \Pi_A(p) {1\choose 1} {1\choose -1}
             + \Pi_F(p) {1\choose -1} {1\choose -1}\, ,
 \\
 \label{eq: KMSpi}
    \Pi_F(p) &=& \Bigl( 1+2n(p_0) \Bigr)\,
                 \Bigl( \Pi_R(p) - \Pi_A(p) \Bigr) \, .
  \end{eqnarray}

One loop results for the polarization tensor are as follows.  We define $N_p =
1+2n(p_0)$ and $D_R(P) = r_P$, $D_A(P) = a_P$, $D_F(P) = f_P$, etc.

\begin{eqnarray}
\Pi_R(K) &=& \frac{1}{2} ig^2 \int\frac{d^4 R}{(2\pi)^4} (a_R f_{K+R} + f_R
r_{K+R}) \\
\Pi_A(K) &=& \frac{1}{2} ig^2  \int\frac{d^4 R}{(2\pi)^4} (r_R f_{K+R} + f_R
a_{K+R})\\
\Pi_F(K) &=& \frac{1}{2} ig^2  \int\frac{d^4 R}{(2\pi)^4} (r_R a_{K+R} + a_R
r_{K+R} + f_R f_{K+R})
\end{eqnarray}

%%%%%%%%%%%%%%%%%%%%%%%%%%%%%%%%%%%%%%%%%%%%%%%%%%%%%%%%%%%%%%%%%%%%%
\section{Three-Point Function}
\label{sec5}
%%%%%%%%%%%%%%%%%%%%%%%%%%%%%%%%%%%%%%%%%%%%%%%%%%%%%%%%%%%%%%%%%%%%%

In the real time formalism, the three-point function has $2^3 = 8$
components. We denote the connected functions
by $\Gamma^C_{abc}$ where $\{a,b,c = 1,2\}$.
In analogy to~(\ref{eq: compD}), they are given by the following
expressions \cite{Chou}:
 \begin{eqnarray}
   \Gamma^C_{111}(x,y,z)
    &=& \langle T(\phi(x)\phi(y)\phi(z))\rangle \, , \nonumber\\
   \Gamma^C_{112}(x,y,z)
    &=& \langle\phi(z)\,T(\phi(x)\phi(y))\rangle \, , \nonumber\\
   \Gamma^C_{121}(x,y,z)
    &=& \langle\phi(y)\,T(\phi(x)\phi(z))\rangle \, , \nonumber\\
   \Gamma^C_{211}(x,y,z)
    &=& \langle\phi(x)\,T(\phi(y)\phi(z))\rangle \, , \nonumber\\
   \Gamma^C_{122}(x,y,z)
    &=& \langle\tilde{T}(\phi(y)\phi(z))\,\phi(x)\rangle\, , \nonumber\\
   \Gamma^C_{212}(x,y,z)
    &=& \langle\tilde{T}(\phi(x)\phi(z))\,\phi(y)\rangle\, , \nonumber\\
   \Gamma^C_{221}(x,y,z)
    &=& \langle\tilde{T}(\phi(x)\phi(y))\,\phi(z)\rangle\, , \nonumber\\
   \Gamma^C_{222}(x,y,z)
    &=& \langle\tilde{T}(\phi(x)\phi(y)\phi(z))\rangle\, .
 \label{eq: compVer}
 \end{eqnarray}
Only seven of these components are independent because of the identity
 \begin{equation}
   \sum_{a=1}^2\sum_{b=1}^2\sum_{c=1}^2
   (-1)^{a+b+c-3} \Gamma^C_{abc} =0
 \label{eq: circVer}
 \end{equation}
which follows in the same way as~(\ref{3}) from $\theta(x) +
\theta(-x) = 1$.The seven combinations that we use are
defined as \cite{Chou}
 \begin{eqnarray}
   \Gamma^C_{R} &=& \Gamma^C_{111} - \Gamma^C_{112}
                - \Gamma^C_{211} + \Gamma^C_{212},
 \nonumber\\
   \Gamma^C_{Ri} &=& \Gamma^C_{111} - \Gamma^C_{112}
                   - \Gamma^C_{121} + \Gamma^C_{122},
 \nonumber\\
   \Gamma^C_{Ro} &=& \Gamma^C_{111} - \Gamma^C_{121}
                   - \Gamma^C_{211} + \Gamma^C_{221},
 \nonumber\\
   \Gamma^C_{F} &=& \Gamma^C_{111} - \Gamma^C_{121}
                + \Gamma^C_{212} - \Gamma^C_{222},
 \nonumber\\
   \Gamma^C_{Fi} &=& \Gamma^C_{111} + \Gamma^C_{122}
                   - \Gamma^C_{211} - \Gamma^C_{222},
 \nonumber\\
   \Gamma^C_{Fo} &=& \Gamma^C_{111} - \Gamma^C_{112}
                   + \Gamma^C_{221} - \Gamma^C_{222},
 \nonumber\\
   \Gamma^C_{E} &=& \Gamma^C_{111} + \Gamma^C_{122}
                + \Gamma^C_{212} + \Gamma^C_{221}.
 \label{eq: physVer}
 \end{eqnarray}
In coordinate space we always label the first leg of the three-point
function by $x$ and call it the ``incoming leg $(i)$", the third leg we
label by $z$ and call it the ``outgoing leg $(o)$", and the second
(middle) leg we label by $y$. Inserting the definitions~(\ref{eq:
compVer}) into~(\ref{eq: physVer}) one finds
 \begin{eqnarray}
   \Gamma^C_{R} &=&
     \theta_{23} \theta_{31} \langle [[\phi_2,\phi_3], \phi_1] \rangle
   + \theta_{21} \theta_{13} \langle [[\phi_2,\phi_1], \phi_3] \rangle \, ,
 \nonumber\\
   \Gamma^C_{Ri} &=&
     \theta_{12} \theta_{23} \langle [[\phi_1,\phi_2], \phi_3] \rangle
   + \theta_{13} \theta_{32} \langle [[\phi_1,\phi_3], \phi_2] \rangle \, ,
 \nonumber\\
   \Gamma^C_{Ro} &=&
     \theta_{32} \theta_{21} \langle [[\phi_3,\phi_2], \phi_1] \rangle
   + \theta_{31} \theta_{12} \langle [[\phi_3,\phi_1], \phi_2] \rangle \, ,
 \nonumber\\
   \Gamma^C_{F} &=&
     \theta_{12} \theta_{23} \langle \{ [\phi_1,\phi_2], \phi_3\} \rangle
   + \theta_{32} \theta_{21} \langle \{ [\phi_3,\phi_2], \phi_1\} \rangle
   + \theta_{12} \theta_{32} \langle [ \{ \phi_3,\phi_1\}, \phi_2] \rangle
   \, ,
 \nonumber\\
   \Gamma^C_{Fi} &=&
     \theta_{21} \theta_{13} \langle \{ [\phi_2,\phi_1], \phi_3\} \rangle
   + \theta_{31} \theta_{12} \langle \{ [\phi_3,\phi_1], \phi_2\} \rangle
   + \theta_{21} \theta_{31} \langle [ \{ \phi_2,\phi_3\}, \phi_1] \rangle
   \, ,
 \nonumber\\
   \Gamma^C_{Fo} &=&
     \theta_{23} \theta_{31} \langle \{ [\phi_2,\phi_3], \phi_1\} \rangle
   + \theta_{13} \theta_{32} \langle \{ [\phi_1,\phi_3], \phi_2\} \rangle
   + \theta_{13} \theta_{23} \langle [ \{ \phi_1,\phi_2\}, \phi_3] \rangle
   \, ,
 \nonumber\\
   \Gamma^C_E &=&
     \theta_{13} \theta_{23} \langle \{ \{\phi_1,\phi_2\}, \phi_3\} \rangle
   + \theta_{21} \theta_{31} \langle \{ \{\phi_2,\phi_3\}, \phi_1\} \rangle
   + \theta_{12} \theta_{32} \langle \{ \{\phi_1,\phi_3\}, \phi_2\} \rangle
   \, ,
 \label{commutators}
 \end{eqnarray}
where we have used the obvious shorthands $\phi_1 \equiv \phi(x)$,
$\phi_2 \equiv \phi(y)$, $\phi_3 \equiv \phi(z)$, and $\theta_{12}
\equiv \theta(x_0-y_0)$, etc. The first three are the retarded three-point
functions; $\Gamma^C_{Ri}$ is retarded with respect to $x_0$,
$\Gamma^C_{Ro}$ is retarded with respect to $z_0$, and $\Gamma^C_R$ is
retarded with respect to $y_0$.

When doing calculations one is usually interested in the 1PI vertex functions which are obtained from the connected functions by truncating external legs.  We will denote 1PI vertex functions by ${\Gamma}$.  
 We can write ${\Gamma}$ as a tensor of the form
\begin{equation}
{\Gamma} = { a \choose b} { c\choose d} {e \choose f}
\end{equation} where the outer product of the column vectors is to be taken. 
For the 1PI functions the analogues of~(\ref{eq: circVer}) and~(\ref{eq: physVer}) are, 
 \begin{equation}
   \sum_{a=1}^2\sum_{b=1}^2\sum_{c=1}^2
    {\Gamma}_{abc} = a+b+c+d+e+f =0,
 \label{eq: circVer2}
 \end{equation}
and,
 \begin{eqnarray}
   {\Gamma}_{R} &=& {\Gamma}_{111} + {\Gamma}_{112}
                + {\Gamma}_{211} + {\Gamma}_{212} \, = \frac{1}{2}(a+b)(c-d)(e+f),
 \nonumber\\
  {\Gamma}_{Ri} &=& {\Gamma}_{111} + {\Gamma}_{112}
                   + {\Gamma}_{121} + {\Gamma}_{122} \, = \frac{1}{2}(a-b)(c+d)(e+f),
 \nonumber\\
   {\Gamma}_{Ro} &=& {\Gamma}_{111} + {\Gamma}_{121}
                   + {\Gamma}_{211} + {\Gamma}_{221} \,=\frac{1}{2}(a+b)(c+d)(e-f) ,
 \nonumber\\
   {\Gamma}_{F} &=& {\Gamma}_{111} + {\Gamma}_{121}
                + {\Gamma}_{212} + {\Gamma}_{222} \, = \frac{1}{2}(a-b)(c+d)(e-f) ,
 \nonumber\\
   {\Gamma}_{Fi} &=& {\Gamma}_{111} + {\Gamma}_{122}
                   + {\Gamma}_{211} + {\Gamma}_{222} \, = \frac{1}{2}(a+b)(c-d)((e-f) ,
 \nonumber\\
   {\Gamma}_{Fo} &=& {\Gamma}_{111} + {\Gamma}_{112}
                   + {\Gamma}_{221} + {\Gamma}_{222} \, = \frac{1}{2}(a-b)(c-d)(e+f) ,
 \nonumber\\
   {\Gamma}_{E} &=& {\Gamma}_{111} + {\Gamma}_{122}
                + {\Gamma}_{212} + {\Gamma}_{221} \, =\frac{1}{2}(a-b)(c-d)(e-f).
 \label{eq: physVer2}
 \end{eqnarray}
The 1PI vertex functions 
${\Gamma}(P_1,P_2,P_3)$ are related to the connected vertex functions $\Gamma(P_1,P_2,P_3)$ as follows: 
\begin{eqnarray}
\Gamma_{R} &=& i^3 a_1 r_2 a_3 {\Gamma}_{R}\nonumber \\
\Gamma_{Ri} &=& i^3 r_1 a_2 a_3 {\Gamma}_{Ri} \nonumber \\
\Gamma_{Ro} &=& i^3 a_1 a_2 r_3 {\Gamma}_{Ro} \nonumber \\
\Gamma_{F} &=& i^3[ r_1 a_2 f_3 {\Gamma}_{Ri}
+  f_1 a_2 r_3 {\Gamma}_{Ro} +  r_1 a_2 r_3 {\Gamma}_{F}] \nonumber \\
\Gamma_{Fi} &=& i^3 [ a_1 r_2 f_3 {\Gamma}_{R} 
+  a_1 f_2 r_3 {\Gamma}_{Ro} 
+ a_1 r_2 r_3 {\Gamma}_{Fi}] \nonumber \\
\Gamma_{Fo} &=& i^3 [ r_1 f_2 a_3 {\Gamma}_{Ri} 
+  f_1 r_2 a_3 {\Gamma}_{R}
+  r_1 r_2 a_3 {\Gamma}_{Fo}] \nonumber \\
{\Gamma}_{E} &=& i^3 [ 
f_1 r_2 f_3 {\Gamma}_{R} +  r_1 f_2 f_3  {\Gamma}_{Ri} +  
f_1 f_2 r_3 {\Gamma}_{Ro} \nonumber \\ 
&& \hspace*{2.2cm} +   
r_1 f_2 r_3 {\Gamma}_{F} +  
f_1 r_2 r_3 {\Gamma}_{Fi} +  
r_1 r_2 f_3 {\Gamma}_{Fo} 
+ r_1 r_2 r_3 {\Gamma}_{E} ] \nonumber 
\end{eqnarray}

For calculational purposes, it is useful to obtain a decomposition of
the 1PI three-point function in terms of the seven functions~(\ref{eq: physVer2}), in analogy to~(\ref{eq: decompD1}) for the two-point
function. Inverting~(\ref{eq: physVer2}) we obtain
 \begin{eqnarray}
   4\,{\Gamma} &=& {\Gamma}_R {1 \choose 1} {1\choose -1} {1\choose 1}
               + {\Gamma}_{Ri} {1 \choose -1}{1\choose 1} {1 \choose 1}
 \nonumber\\
           &+& {\Gamma}_{Ro} {1 \choose 1} {1\choose 1} {1 \choose -1}
             + {\Gamma}_F {1\choose -1} {1\choose 1}{1\choose -1}
 \nonumber\\
           &+& {\Gamma}_{Fi} {1\choose 1}{1\choose -1}{1\choose -1}
             + {\Gamma}_{Fo} {1\choose -1}{1\choose -1}{1\choose 1}
 \nonumber\\
           &+& {\Gamma}_E {1\choose -1}{1\choose -1}{1\choose -1}.
 \label{eq: decompver}
 \end{eqnarray}

One loop expressions for the three point functions are as follows.  We use 
$P_-= P-K/2$, $P_+ = -(P+K/2)$, $P_1 = P+R-K/2$ and $P_3 = -(P+R+K/2)$ (see
Fig.~\ref{fig1}).
\begin{eqnarray}
{\Gamma}_R (P_-,K,P_+) &=& \frac{1}{2}g^3\int \frac{d^4 R}{(2\pi)^4} (f_R a_1
a_3 + r_R f_1 a_3 + a_R a_1 f_3) \nonumber\\
{\Gamma}_{Ri} (P_-,K,P_+) &=& \frac{1}{2}g^3\int \frac{d^4 R}{(2\pi)^4} (f_R
r_1 a_3 + a_R f_1 r_3 + a_R r_1 f_3) \nonumber \\
{\Gamma}_{Ro} (P_-,K,P_+) &=& \frac{1}{2}g^3\int \frac{d^4 R}{(2\pi)^4} (f_R
a_1 r_3 + r_R f_1 r_3 + r_R r_1 f_3) \label{Gint}\\
{\Gamma}_F (P_-,K,P_+) &=& \frac{1}{2}g^3\int \frac{d^4 R}{(2\pi)^4} (f_R(
f_1 r_3 + f_3 r_1) + a_R a_3 r_1 + r_R r_3 a_1) \nonumber\\
{\Gamma}_{Fi} (P_-,K,P_+) &=& \frac{1}{2}g^3\int \frac{d^4 R}{(2\pi)^4}
(f_3(r_R f_1 + f_R a_1) + r_R r_1 r_3 + a_R a_1 a_3 ) \nonumber\\
{\Gamma}_{Fo} (P_-,K,P_+) &=& \frac{1}{2}g^3\int \frac{d^4 R}{(2\pi)^4}
(f_1(f_R a_3 + a_R f_3) + r_R a_3 a_1 + a_R r_3 r_1)\nonumber
\end{eqnarray}

We are interested in the infra-red limit $K\rightarrow 0$.  We use the splitting
identity of the form
\begin{eqnarray}
D_1 D_3 &=& \frac{1}{P_3^2-P_1^2}(D_1-D_3) = \frac{1}{2K\cdot (P+R)}(D_1 - D_3)
\approx  \frac{1}{2K\cdot P}(D_1 - D_3)\label{eq: splitD} 
\end{eqnarray}
where we take the limit $R\rightarrow 0$ to study the infra-red limit.  This
identity allows us to rewrite the one-loop three point vertex in terms of a
difference in one-loop two point functions.  We will work with the vertex function that is retarded with respect to the middle leg and obtain, 
\begin{eqnarray}
\label{eq: xx}
\Gamma_R(P_-,K,P_+) &=& -\frac{ig}{2P\cdot K} ( \Pi_A(P_-) - \Pi_A(P_+)),
\end{eqnarray}
which is another form of~(\ref{relation}) that is valid at finite temperature.  

%%%%%%%%%%%%%%%%%%%%%%%%%%%%%%%%%%%%%%%%%%%%%%%%%%%%%%%%%%%%%%%%%%%%%%%%%

%%%%%%%%%%%%%%%%%%%%%%%%%%%%%%%%%%%%%%%%%%%%%%%%%%%%%%%%%%%%%%%%%%%%%
\section{Four-Point Function}
\label{sec6}
%%%%%%%%%%%%%%%%%%%%%%%%%%%%%%%%%%%%%%%%%%%%%%%%%%%%%%%%%%%%%%%%%%%%%
The 1PI four-point function forms a 16 component tensor which we can write as the
outer product of four two component vectors,
\begin{equation}
M = {a \choose b} {c\choose d} {e\choose f} {g\choose h}
\end{equation}

Then the retarded 1PI four-point functions are given by
\begin{eqnarray}
M_{R1} &=& M_{1111} + M_{1112} + M_{1121} + M_{1211} + M_{1122} + M_{1212} +
M_{1221} + M_{1222} \nonumber\\
   &=& \frac{1}{2} (a-b)(c+d)(e+f)(g+h)\nonumber \\
M_{R2} &=& M_{1111} + M_{1112} + M_{1121} + M_{2111} + M_{1122} + M_{2112} +
M_{2121} + M_{2122}\nonumber \\
   &=& \frac{1}{2} (a+b)(c-d)(e+f)(g+h)\\
M_{R3} &=& M_{1111} + M_{1112} + M_{2111} + M_{1211} + M_{2112} + M_{1212} +
M_{2211} + M_{2212} \nonumber\\
   &=& \frac{1}{2} (a+b)(c+d)(e-f)(g+h)\nonumber\\
M_{R4} &=& M_{1111} + M_{2111} + M_{1121} + M_{1211} + M_{2121} + M_{2211} +
M_{1221} + M_{2221} \nonumber\\
   &=& \frac{1}{2} (a+b)(c+d)(e+f)(g-h)\nonumber
\end{eqnarray}
where we have used the relation
\begin{equation}
\sum_{a,b,c,d=1}^{2} M_{abcd} = 0.
\end{equation}
The other combinations we will need are,
\begin{eqnarray}
M_A &=& \frac{1}{2}(a+b) (c+d) (e-f) (g-h)\nonumber \\
M_B &=& \frac{1}{2}(a-b) (c+d) (e-f) (g+h)\nonumber\\
M_C &=& \frac{1}{2}(a+b) (c-d) (e-f) (g+h) \\
M_D &=& \frac{1}{2}(a+b) (c-d) (e+f) (g-h) \nonumber\\
M_E &=& \frac{1}{2}(a-b) (c-d) (e+f) (g+h) \nonumber\\
M_F &=& \frac{1}{2}(a-b) (c+d) (e+f) (g-h) \nonumber
\end{eqnarray}

The one loop results are,
\begin{eqnarray}
M_{R1} &=& \frac{1}{2} g^4 \int \frac{d^4R}{(2\pi)^4} ( f_R a_1 r_3 a_4 + r_R
f_1 r_3 a_4 + r_R r_1 f_3 a_4 + a_R a_1 r_3 f_4 )\nonumber\\
M_{R2} &=& \frac{1}{2} g^4 \int \frac{d^4R}{(2\pi)^4} ( f_R a_1 a_3 a_4 + r_R
f_1 a_3 a_4 + a_R a_1 f_3 r_4 + a_R a_1 a_3 f_4 )\nonumber\\
M_{R3} &=& \frac{1}{2} g^4 \int \frac{d^4R}{(2\pi)^4} ( f_R r_1 a_3 a_4 + a_R
f_1 r_3 r_4 + a_R r_1 f_3 r_4 + a_R r_1 a_3 f_4 )\nonumber\\
M_{R4} &=& \frac{1}{2} g^4 \int \frac{d^4R}{(2\pi)^4} ( f_R a_1 r_3 r_4 + r_R
f_1 r_3 r_4 + r_R r_1 f_3 r_4 + r_R r_1 a_3 f_4 )\nonumber\\
M_{A} &=& \frac{1}{2} g^4 \int \frac{d^4R}{(2\pi)^4} ( r_R a_1 r_3 r_4 + a_R
r_1 a_3 a_4 + f_R(f_1 r_3 r_4 + r_1 f_3 r_4 + r_1 a_3 f_4)) \\
M_{B} &=& \frac{1}{2} g^4 \int \frac{d^4R}{(2\pi)^4} ( r_R a_1 r_3 a_4 + a_R
r_1 a_3 r_4 + f_R f_1 r_3 a_4 + a_R f_1 r_3 f_4 + f_R r_1 f_3 a_4 + a_R r_1 f_3
f_4 )\nonumber\\
M_{C} &=& \frac{1}{2} g^4 \int \frac{d^4R}{(2\pi)^4} ( a_R r_1 r_3 r_4 + r_R
a_1 a_3 a_4 + f_1(a_R f_3 r_4 + f_R a_3 a_4 + a_R a_3 f_4) ) \nonumber\\
M_{D} &=& \frac{1}{2} g^4 \int \frac{d^4R}{(2\pi)^4} ( r_R r_1 r_3 r_4 + a_R
a_1 a_3 a_4 + r_R f_1 f_3 r_4 + f_R a_1 f_3 r_4 + r_R f_1 a_3 f_4 + f_R a_1 a_3
f_4)\nonumber\\
M_{E} &=& \frac{1}{2} g^4 \int \frac{d^4R}{(2\pi)^4} ( r_R r_1 r_3 a_4 + a_R
a_1 a_3 r_4 + f_3(r_R f_1 a_4 + f_R a_1 a_4 + a_R a_1 f_4)) \nonumber\\
M_{F} &=& \frac{1}{2} g^4 \int \frac{d^4R}{(2\pi)^4} ( a_R a_1 r_3 a_4 + r_R
r_1 a_3 r_4 + f_4(f_R a_1 r_3 + r_R f_1 r_3 + r_R r_1 f_3) f_4 ) \label{Ms}
\end{eqnarray}
Variables are defined as in Fig.~\ref{fig2}.

Using~(\ref{Gint}) and~(\ref{eq: splitD}) we obtain the splitting relations that we will need:  
\begin{eqnarray}
M_{R2} (P_-,K,P_3,R) &=& \frac{g}{2P\cdot K} [\Gamma_R(P_-,-P_1,R) -
\Gamma_{Ri}(-P_+, P_3,R)] \label{splitM}\\
M_{R2}(P_1,K,P_+,-R) &=& \frac{g}{2P\cdot K} [ \Gamma_R(P_1,-P_-,-R) +
\Gamma_R^*(-P_+,P_3,R)]\nonumber
\end{eqnarray}

%%%%%%%%%%%%%%%%%%%%%%%%%%%%%%%%%%%%%%%%%%%%%%%%%%%%%%%%%%%%%%%%%%%%%
\section{Five-Point Function}
\label{sec7}
%%%%%%%%%%%%%%%%%%%%%%%%%%%%%%%%%%%%%%%%%%%%%%%%%%%%%%%%%%%%%%%%%%%%%
The 1PI five-point function forms a 32 component tensor which we can write as the
outer product of five two component vectors,
\begin{equation}
M = {a \choose b} {c\choose d} {e\choose f} {g\choose h} {i\choose j}
\end{equation}

The vertices that we need are given by,
\begin{eqnarray}
C_{R2} &=& \frac{1}{2} (a+b)(c-d)(e+f)(g+h)(i+j)\nonumber \\
C_{\alpha} &=&\frac{1}{2} (a+b)(c-d)(e+f)(g-h)(i+j)\nonumber \\
C_{\beta} &=&\frac{1}{2} (a+b)(c-d)(e+f)(g+h)(i-j)\nonumber 
\end{eqnarray}
The one loop results are given by,
\begin{eqnarray}
C_{R2} &=& \frac{1}{2} g^5 \int \frac{d^4L}{(2\pi)^4} [r_L r_L f_1 a_3 r_{R+L} + a_L r_L a_1 a_3 f_{R+L} + a_L f_L a_1 a_3 a_{R+L} + a_L a_L a_1 f_3 a_{R+L} \nonumber\\
&&\hspace*{2cm} + r_L f_L a_1 a_3 r_{R+L} ]\nonumber \\
C_{\alpha} &=& \frac{1}{2} g^5 \int \frac{d^4L}{(2\pi)^4} [r_L a_L r_1 r_3 r_{R+L} + a_L r_L f_1 f_3 r_{R+L} + r_L f_L f_1 a_3 r_{R+L} + a_L r_L a_1 a_3 a_{R+L} \nonumber \\
&&\hspace*{2cm} + a_L f_L a_1 a_3 f_{R+L} + a_l a_l a_1 f_3 f_{R+L} + f_l f_l a_1 a_3 r_{R+L} + a_L f_L a_1 f_3 r_{R+L} ]\nonumber \\
C_{\beta} &=& \frac{1}{2} g^5 \int \frac{d^4L}{(2\pi)^4} [ r_L a_L r_1 r_3 a_{R+L} + r_L r_L f_1 a_3 f_{R+L} + r_L a_L f_1 f_3 a_{R+L} + r_L f_L f_1 a_3 a_{R+L} \nonumber\\
&&\hspace*{2cm}+ r_L a_L a_1 a_3 r_{R+L} + r_L f_L a_1 a_3 f_{R+L} + f_L f_L a_1 a_3 a_{R+L} + a_L f_L a_1 f_3 a_{R+L} ]\nonumber 
\end{eqnarray}
Variables are defined as in Fig.~\ref{fig2a}.

Using~(\ref{eq: splitD}) and~(\ref{Ms}) we obtain, in the infra-red limit, the relations,
\begin{eqnarray}
C_{R2}(P_-,K,P_+,R,-R) &=& -\frac{g}{2P\cdot K}[M_{R2}^*(-P_-,P_-,-R,R) +
M_{R2}(P_+,-P_+,-R,R)]\nonumber \\
C_{\alpha}(P_-,K,P_+,R,-R) &=& \frac{g}{2P\cdot K}[M^*_C(-P_-,P_-,-R,R) -
M_D(P_+,-P_+,-R,R)]\nonumber\\
C_{\beta}(P_-,K,P_+,R,-R) &=& \frac{g}{2P\cdot K}[M^*_D(-P_-,P_-,-R,R) -
M_C(P_+,-P_+,-R,R)]\label{Csplit}
\end{eqnarray}
%%%%%%%%%%%%%%%%%%%%%%%%%%%%%%%%%%%%%%%%%%%%%%%%%%%%%%%%%%%%%%%%
\section{Ladder resummation and cancellation at finite temperature}
\label{sec9}
%%%%%%%%%%%%%%%%%%%%%%%%%%%%%%%%%%%%%%%%%%%%%%%%%%%%%%%%%%%%%%%%%%%%%%%%%%
We earlier discussed at zero temperature how the full Schwinger--Dyson
equation of Fig.~\ref{fullselfenergy} simplifies to  Eq.~(\ref{simplify})
when the relation of Eq.~(\ref{relation}) holds between the vertex
and self--energy function. In this section we generalize this
result to finite temperature. In
this formalism the full Schwinger--Dyson equation has the form
\begin{eqnarray}
\Pi_R(K) = \frac{1}{2}ig^2 \int \frac{d^4 R}{(2\pi)^4} &[& \Delta_A(R)
\Delta_F(K+R) + \Delta_F(R) \Delta_R(K+R) \nonumber \\
 &+& i\{-\Gamma^*_{Ri}(-K-R,K,R) \Delta_F(R) \Delta_R(K+R) \nonumber \\
&-& \Gamma^*_{Ro}(-K-R,K,R) \Delta_A(R) \Delta_F(K+R)  \label{int}\\
&+& \Gamma^*_F(-K-R,K,R) \Delta_A(R) \Delta_R(K+R) \}]\nonumber
\end{eqnarray}
where the propagators are full ones.
We rewrite this expression by using the splitting relations for the three point
vertices, analogous to Eq.~(\ref{eq: xx}). We obtain
\begin{eqnarray}
\Gamma^*_{Ri}(-K-R,K,R) &=& -\frac{ig}{K^2+2K\cdot R}(\Pi_R(K+R) - \Pi_R(R))
\nonumber \\
\Gamma^*_{Ro}(-K-R,K,R) &=& -\frac{ig}{K^2+2K\cdot R}(\Pi_A(K+R) - \Pi_A(R))
\label{sv} \\
\Gamma^*_{F}(-K-R,K,R) &=& \frac{ig}{K^2+2K\cdot R}(\Pi_F(K+R) - \Pi_F(R))
\nonumber
\end{eqnarray}
Splitting the propagators gives
\begin{eqnarray}
\Delta_R(R) \Delta_R(K+R)  &=& \frac{\Delta_R(R) - \Delta_R(K+R) }{K^2+ 2K\cdot
R + \Pi_R(R) - \Pi_R(K+R)} \nonumber \\
\Delta_A(R) \Delta_R(K+R)  &=& \frac{\Delta_A(R) - \Delta_R(K+R) }{K^2+ 2K\cdot
R + \Pi_A(R) - \Pi_R(K+R)} \label{sp}
\end{eqnarray}
and substituting Eq.~(\ref{sv}) and Eq.~(\ref{sp}) into Eq.~(\ref{int}) 
we obtain
\begin{eqnarray}
\Pi_R(K) = ig^2 \int \frac{d^4 R}{(2\pi)^4} \frac{\Delta_F(R) }{K^2 + 2K\cdot
R},
\label{Erel}
\end{eqnarray}
which is the desired result.
\par
As a specific example of this type of simplification, we consider the set of
diagrams shown in Figs.~\ref{oneloop}, \ref{firstladder}, and
\ref{cancel}, which we call respectively
 $\Pi^{(1)}$,  $\Pi^{(2)}$, and  $\Pi^{(3)}$.
 The diagram of Fig.~(\ref{oneloop})
is given after splitting the propagators as
\begin{eqnarray}
\Pi_R ^{(1)}(K) = i g^2 \int \frac{d^4 R}{(2\pi)^4} \frac{f_R}{k^2 + 2 K\cdot
R}
\label{pi1}
\end{eqnarray}
The diagram of Fig.~\ref{cancel} is given by the integral
\begin{eqnarray}
\Pi_R^{(2)}(K) &=& \frac{1}{4}g^4 \int \frac{d^4 R}{(2\pi)^4} \int \frac{d^4
R'}{(2\pi)^4}\nonumber \\
&[&
r_{R+K}a_R(a_{R+R'}f_R  f_{R'} + a_{R+R'} r_R r_{R'} + r_{R+R'} r_R a_{R'} +
f_{R+R'} r_R f_{R'} + f_{R+R'} f_R r_{R'}) \nonumber \\
&+& r_{R+K} f_R (r_{R+R'} r_R f_{R'} + r_{R+R'} f_R r_{R'} + a_{R+R'} f_R
a_{R'} + f_{R+R'} r_R a_{R'} )\\
&+& f_{R+K} a_R ( a_{R+R'} a_R f_{R'} + f_{R+R'} a_R r_{R'}) + f_{R+K} f_R
(r_{R+R'} a_R r_{R'} + a_{R+R'} a_R a_{R'}) \nonumber \\
&+& a_{R+K} r_R ( r_{R+R'} a_R r_{R'} + a_{R+R'} a_R a_{R'})]\nonumber
\end{eqnarray}
We split pairs of propagators of the form $D_R D_{R+K}$ and rewrite the result
by extracting polarization tensors from the factors involving integrals over
$R'$.  There are two pieces.  The first does not involve propagators that
depend on $R+K$ and is given by,
\begin{eqnarray}
&&\frac{ig^2}{2} \int \frac{d^4 R}{(2\pi)^4} \frac{1}{K^2+2K\cdot R} [a_R f_R
\Pi_A(R) + a_R r_R \Pi_F(R) + f_R r_R \Pi_R(R)] \nonumber \\
&=& \frac{ig^2}{2} \int \frac{d^4 R}{(2\pi)^4} \frac{1}{K^2+2K\cdot R} N_R
(r_R^2 \Pi_R(R) - a_R^2 \Pi_A(R))
\label{pi2a}
\end{eqnarray}
Multiplying this result by two, as indicated by the Schwinger-Dyson equation,
and combining with the zeroth order result (\ref{pi1}) gives,
\begin{eqnarray}
ig^2 \int \frac{d^4 R}{(2\pi)^4} \frac{1}{K^2+2K\cdot R} N_R [ r_R + r_R^2
\Pi_R(R) - (a_R + a_R^2 \Pi_R(R))]
\end{eqnarray}
which is the first two terms in the expansion of Eq.~(\ref{Erel}).
\par
Next we show that the second piece of $\Pi_R^{(2)}$ cancels with $\Pi_R^{(3)}$.
 The piece of $\Pi_R^{(2)}$ that contains the propagator $D_{R+K}$ is given by
(with the factor of two from the Schwinger-Dyson equation),
\begin{eqnarray}
-ig^2  \int \frac{d^4 R}{(2\pi)^4} \frac{1}{K^2 + 2 K\cdot R} (r_{R+K} a_R
\Pi_F(R) + r_{R+K} f_R \Pi_R(R) + f_{R+K} a_R \Pi_A(R))
\label{pi2b}
\end{eqnarray}
We compare this result with the expression for the diagram
Fig.~\ref{firstladder},
\begin{eqnarray}
\Pi_R^{(3)} &=& -\frac{1}{4} g^4  \int \frac{d^4 R}{(2\pi)^4}  \int \frac{d^4
R'}{(2\pi)^4} \nonumber \\
&[& a_R r_{R+K} ( r_{K+R+R'} r_{R+R'} a_{R'} + r_{K+R+R'} f_{R+R'}f_{R'} +
f_{K+R+R'} a_{R+R'} f_{R'} + a_{K+R+R'} r_{R'} a_{R+R'}) \nonumber \\
&+& a_R f_{R+K} ( a_{K+R+R'} a_{R+R'} f_{R'} + r_{K+R+R'} f_{R+R'} r_{R'} +
f_{K+R+R'} a_{R+R'} r_{R'}) \\
&+& f_R r_{R+K} (r_{K+R+R'} r_{R+R'} f_{R'} + r_{K+R+R'} f_{R+R'} a_{R'} +
f_{K+R+R'} a_{R+R'} a_{R'})]\nonumber
\end{eqnarray}
We do the variable shift $R\rightarrow -(R+K)$ on the second term, and split
pairs of propagators of the form $D_{K+R+R'} D_{R+R'}$.  This splitting
introduces a denominator $K^2 + 2K\cdot(R+R')$.  We take the infrared limit
$K^2 + 2K\cdot(R+R') \rightarrow K^2 + 2K\cdot R$ and extract polarization
tensors as before to obtain,
\begin{eqnarray}
ig^2  \int \frac{d^4 R}{(2\pi)^4} \frac{1}{K^2+2K\cdot R} (f_R r_{R+K} \Pi_R(R)
+ f_{R+K} a_R \Pi_A(R) + a_R r_{R+K} \Pi_F(R))
\label{pi3}
\end{eqnarray}
which cancels with Eq.~(\ref{pi2b}).
%%%%%%%%%%%%%%%%%%%%%%%%%%%%%%%%%%%%%%%%%%%%%%%%%%%%%%%%%%%%%%%%
\section{Ladder resummations}
\label{sec10}
%%%%%%%%%%%%%%%%%%%%%%%%%%%%%%%%%%%%%%%%%%%%%%%%%%%%%%%%%%%%%%%%%%%%%%%%%%
In this section we show in the infrared limit that a wide class of
vertices and polarization tensors (Figs.~\ref{fig3}--\ref{fig10})
with internal vertex and self energy corrections
satisfy
Eq.~(\ref{eq: xx}). As discussed earlier, the significance of this
result is twofold. First of all, it allows us one construct an expansion
of ladder graphs in terms of the iterative solution to a partial
Schwinger--Dyson equation. Secondly, it can be used to show that
the solution
to the Schwinger--Dyson equation for the full self--energy simplifies
considerably when these ladder graphs are combined with 
certain non--ladder graphs.
\par
  We start with Fig.~\ref{fig3}.  The vertex is given by
\begin{equation}
\Gamma_{abc}(P_-,K,P_+) = i^4 (-ig)^3 \int \frac{d^4R}{(2\pi)^4} (\tau_3
D(R))_{cd} (-i\Pi_{de}(R))(D(R) \tau_3)_{ea} D_{ab}(P_1) (D(P_3)\tau_3)_{cb}
\end{equation}
We use~(\ref{eq: decompD1}) and~({\ref{eq: Pidecomp1}) to write the 
propagators and polarization tensors as
the outer products of column vectors and expand indices according to the rules
given in \cite{PeterH}, \cite{Carr}. We take only those terms that we
need to get $\Gamma_R$~(\ref{eq: physVer}). The result is,
\begin{eqnarray}
\Gamma_R(P_-,K,P_+) = \frac{1}{2} g^3  \int\frac{d^4 R}{(2\pi)^4} [\Pi_R(R) 
r_R a_3
(r_R f_1 + f_R a_1) &+& \Pi_A(R) a_R a_1(f_R a_3+a_R f_3)\nonumber \\
& + &
\Pi_F(R) a_R r_R a_1 a_3 ]\nonumber
\end{eqnarray}
We split products of propagators of the form $S_1 S_3$ to obtain,
\begin{eqnarray}
\Gamma_R(P_-,K,P_+) &=& \frac{1}{2} g^3 \frac{1}{2P\cdot K} \int\frac{d^4
R}{(2\pi)^4}
\{ \Pi_R(R)[r_R^2 f_1  + r_R f_R (a_1 - a_3)]\nonumber \\
&+& \Pi_A [a_R f_R (a_1 - a_3) - a_R^2  f_3]
+ \Pi_F a_R r_R (a_1 - a_3) \}
\label{eq: aaa}
\end{eqnarray}
 To rewrite this equation we compare with the expression for the
polarization tensor in Fig.~\ref{fig4}.  The result is,
\begin{eqnarray}
\Pi_R(K) = \frac{1}{2} ig^2 \int\frac{d^4 R}{(2\pi)^4} [\Pi_A(R) (r_{R+K} a_R
f_R + f_{R+K} a_R a_R) &+& \Pi_R(R) r_{R+K} f_R r_R \nonumber \\
&+& \Pi_F(R) r_{R+K} a_R r_R ]
\label{eq: bbb}
\end{eqnarray}
Comparison shows that~(\ref{eq: aaa}),~(\ref{eq: bbb}) satisfy~(\ref{eq: xx})
as required.

Next we consider Fig.~\ref{fig5}.
The three vertex diagrams give respectively,
\begin{eqnarray}
\Gamma_R(P_-,K,P_+) &=& \frac{i}{2} g^2 \int\frac{d^4 R}{(2\pi)^4}
[\Gamma_{Fi}(P_1,K,P_3)a_1 r_3 a_R + \Gamma_{Fo}(P_1,K,P_3)r_1 a_3 r_R
\nonumber\\
& & \hspace*{2cm}+ \Gamma_R(P_1,K,,P_3)(a_1 a_R f_3 + a_1 f_R a_3 + f_1 r_R
a_3)]\\
 &+& \frac{1}{2} g^3 \int\frac{d^4 R}{(2\pi)^4}
[\Pi_R(P_1) r_1 f_1 a_3 r_R + \Pi_F(P_1)r_1 a_1 r_R a_3 \nonumber\\
& & \hspace*{2cm} + \Pi_A(P_1) a_1 (a_3 r_R f_1 + a_3 f_R a_1 + f_3 a_R
a_1)]\nonumber\\
 &+& \frac{1}{2} g^3 \int\frac{d^4 R}{(2\pi)^4}
[\Pi_R(P_3) f_3 r_3 a_R a_1 + \Pi_F(P_3) a_3 r_3 a_1 a_R \nonumber\\
& & \hspace*{2cm}  + \Pi_A(P_3) a_3(a_1 a_R f_3  + a_1 f_R a_3 + f_1 r_R
a_3)]\nonumber
\end{eqnarray}
We rewrite this expression as follows.  We note that the terms proportional to
$\Gamma_{Fi}$ and $\Gamma_{Fo}$ give zero after integration (assuming no poles
in the vertices themselves).  We rewrite the term proportional to $\Gamma_R$ by
using the relation ~(\ref{eq: xx}) that we are trying to verify.
 Finally, for the terms that contain a polarization
tensor, we split products of propagators of the form $S_1 S_3$.  We give the result below.
\begin{eqnarray}
\Gamma_R(P_-,K,P_+) &=& \frac{1}{2} g^3 \frac{1}{2P\cdot K} \int \frac{d^4
R}{(2\pi)^4}
[\Pi_A(P_1)(r_R a_1 f_1 + f_R a_1 a_1) - \Pi_A(P_3)(a_R a_3 f_3 + f_R a_3 a_3)
\nonumber\\
&+& \Pi_R(P_1) r_R r_1 f_1 - \Pi_R(P_3) a_R r_3 f_3 + \Pi_F(P_1) r_R r_1 a_1 -
\Pi_F(P_3) a_R r_3 a_3 ]
\end{eqnarray}

The expression for the polarization tensor of Fig.~\ref{fig6} is,
\begin{equation}\Pi_A(P_-) = \frac{1}{2} ig^2 \int\frac{d^4 R}{(2\pi)^4}
[\Pi_F(P_1)r_R r_1 a_1 + \Pi_R(P_1) r_R r_1 f_1 + \Pi_A(P_1) a_1(r_R f_1  + f_R
a_1)]
\end{equation}
Comparison shows that~(\ref{eq: xx}) is satisfied.

Next we consider the vertices in figure Fig.~\ref{fig7}.
{}From left to right, top to bottom, we obtain for the first diagram
\begin{eqnarray}
 \Gamma_R(P_-,K,P_+) &=& \frac{i}{2} g^2 \int\frac{d^4 R}{(2\pi)^4}
[\Gamma_{Ri}(-P_1,P_-,R)a_1 (a_R f_3+ f_R a_3) + 
\Gamma_{Ro}(-P_1,P_-,R) r_R f_1
a_3\nonumber \\
 &+& \Gamma_F (-P_1,P_-,R) r_R a_1 a_3 ]
\end{eqnarray}
Splitting the propagators of the form $S_1 S_3$ using~(\ref{eq: splitD}) we have,
\begin{eqnarray}
 \Gamma_R(P_-,K,P_+) &=& \frac{i}{2} g^2\frac{1}{2P\cdot K} \int\frac{d^4
R}{(2\pi)^4} [\Gamma_{Ri}(-P_1,P_-,R)(f_R (a_1-a_3) - a_R f_3) \nonumber\\
&+&\Gamma_{Ro}(-P_1,P_-,R) r_R f_1 + \Gamma_F (-P_1,P_-,R) r_R a_1 ]
\end{eqnarray}
The second diagram yields
\begin{eqnarray}
 \Gamma_R(P_-,K,P_+) &=& \frac{i}{2} g^2 \int\frac{d^4 R}{(2\pi)^4}
[\Gamma_R(P_+,-P_3, -R)a_3(r_R f_1 + f_R a_1) \nonumber\\
&+& \Gamma_{Ro}(P_+,-P_3, -R) a_R a_1 f_3 + \Gamma_{Fi}(P_+,-P_3, -R) a_R a_1
a_3]
\end{eqnarray}
and splitting propagators we obtain,
\begin{eqnarray}
 \Gamma_R(P_-,K,P_+) &=& \frac{i}{2} g^2\frac{1}{2P\cdot K} \int\frac{d^4
R}{(2\pi)^4} [\Gamma_R(P_+,-P_3, -R)(r_R f_1 + f_R (a_1 - a_3))\nonumber\\
&-&  \Gamma_{Ro}(P_+,-P_3, -R)a_R f_3 - \Gamma_{Fi}(P_+,-P_3, -R) a_R a_3]
\end{eqnarray}
The third diagram of Fig.~\ref{fig7} gives
\begin{eqnarray}
 \Gamma_R(P_-,K,P_+) &=& \frac{i}{2} g  \int\frac{d^4 R}{(2\pi)^4} [
M_{R2}(P_-,K,P_3,R)(f_R a_3+ f_3 a_R) \nonumber\\
&+& M_C (P_-,K,P_3,R)a_R r_3 + M_D (P_-,K,P_3,R)r_R a_3 ]
\end{eqnarray}
We drop the last two terms since they give zero by contour integration and
rewrite the first term using the relation~(\ref{splitM})
which gives,
\begin{equation}
 \Gamma_R(P_-,K,P_+) = \frac{i}{2}g^2 \frac{1}{2P\cdot K}  \int\frac{d^4
R}{(2\pi)^4}
[\Gamma_R(P_-,-P_1,R) - \Gamma_{Ri}(-P_+, P_3,R)](f_R a_3 + f_3 a_R)
\end{equation}
Finally, the fourth diagram of Fig.~\ref{fig7} yields
\begin{eqnarray}
 \Gamma_R(P_-,K,P_+) & =& \frac{i}{2} g  \int\frac{d^4 R}{(2\pi)^4} [
M_{R2}(P_1,K,P_+,-R)(f_R a_1 + f_1 r_R)\nonumber\\ 
&+& M_C(P_1,K,P_+,-R)r_R r_1
+M_D(P_1,K,P_+,-R)a_R a_1].
\end{eqnarray}
Dropping the terms that give zero after integration and using the relation~(\ref{splitM}) 
gives,
\begin{equation}
 \Gamma_R(P_-,K,P_+) = \frac{i}{2}g^2 \frac{1}{2P\cdot K}  \int\frac{d^4
R}{(2\pi)^4}
 [ \Gamma_R(P_1,-P_-,-R) + \Gamma_R^*(-P_+,P_3,R)](f_R a_1+ f_1 r_R)
\end{equation}

We need to compare with the expressions for the polarization tensors given
in Fig.~\ref{fig8}.  We have,
\begin{eqnarray}
\Pi_A(P_-) &=& -\frac{1}{2} g \int\frac{d^4 R}{(2\pi)^4} [
\Gamma_R(P_1,-P_-,-R) ( f_R a_1 + r_R f_1) + \Gamma_{Ri}(-P_1,P_-,R) f_R a_1
\nonumber \\
& &\hspace*{2cm} + \Gamma_{Ro} (-P_1,P_-,R)r_R f_1 + \Gamma_F(-P_1,P_-,R) r_R
a_1 ]\\
\Pi_A(P_+) &=&  -\frac{1}{2} g \int\frac{d^4 R}{(2\pi)^4} [\Gamma_{Ri}(-P_+,
P_3,R)(f_R a_3 + a_R f_3) + \Gamma_R(P_+,-P_3,-R) f_R a_3  \nonumber\\
& & \hspace*{2cm}+ \Gamma_{Ro}(P_+,-P_3,-R) a_R f_3 + \Gamma_{Fi}(P_+,-P_3,-R)
a_R a_3 ]
\end{eqnarray}
Using $\Gamma_{Rx}(\{P_i\}) = -\Gamma_{Rx}^*(-\{P_i\})$ and 
$\Gamma_{Fx}(\{P_i\}) = 
\Gamma_{Fx}^*(\{-P_i\})$, where $\{x,i\}=\{1,2,3\}$,
we find that Eq.~(\ref{eq: xx}) is satisfied.

Lastly we look at Fig.~\ref{fig9}, which is given by
\begin{eqnarray}
 \Gamma_R(P_-,K,P_+) &=& \frac{i}{2}\int\frac{d^4
R}{(2\pi)^4}[C_{R2}(P_-,K,P_+,R,-R) f_R + C_{\alpha}(P_-,K,P_+,R,-R) r_R
\nonumber \\
&+& C_{\beta}(P_-,K,P_+,R,-R) a_R].
\end{eqnarray}
We rewrite this result by using~(\ref{Csplit}).  
We need to compare with the expression for the polarization tensor of
Fig.~\ref{fig10}.
\begin{eqnarray}
\Pi_A(P_-) &=& \frac{i}{2}\int\frac{d^4 R}{(2\pi)^4}[M_{R2} (P_-, -P_-, R, -R)
f_R + M_C (P_-, -P_-, R, -R) r_R\nonumber\\
&+& M_D (P_-, -P_-, R, -R) a_R ]
\end{eqnarray}
Using $M_N(\{P_i\}) = M^*_N(-\{P_i\})$ and $M_{Rx}(\{P_i\}) = 
-M^*_{Rx}(\{P_i\})$, with $N=\{A,B,C,D,E,F\}$ and $\{x,i\}=\{1,2,3,4\}$ 
we find that~(\ref{eq: xx}) is
satisfied.

%Finally, we show that Fig.~\ref{fig11} gives no contribution.  
%This integral is
%\begin{eqnarray}
%\Gamma_R(P_-,K,P_+,) &=& \frac{i}{2} g  \int\frac{d^4
%R}{(2\pi)^4}[M_{R3}(P_-,P_+,R+K,-R) f_R r_{R+K} \nonumber \\
%&+&M_{R4} (P_-,P_+,R+K,-R) a_R f_{R+K} + M_A(P_-,P_+,R+K,-R) a_R r_{R+K} ]
%\end{eqnarray}
%which gives zero when we make use of the identity (????????????)
%\begin{equation}
%M_A(P_a,P_b,P_c,P_d) + N_c M_{R4}(P_a,P_b,P_c,P_d) + N_d
%M_{R3}(P_a,P_b,P_c,P_d)
%\end{equation}
%%%%%%%%%%%%%%%%%%%%%%%%%%%%%%%%%%%%%%%%%%%%%%%%%%%%%%%%%%%%%%%%%%
\section{Conclusions}
We have shown at finite temperature in the infrared
limit that a wide class of vertex and
self--energy graphs with effective vertices and corrected internal propagators
satisfy a simple relation like Eq.~(\ref{eq: xx}).
This result has two consequences. The first is that an effective expansion
of ladder graphs can be realized as the iteration
of a partial Schwinger--Dyson equation (Fig.~\ref{semiselfenergy}). 
The second is that these ladder graphs may be combined
with certain non--ladder graphs in 
the Schwinger--Dyson equation
(Fig.~\ref{fullselfenergy})
to give a relatively simple expression for the full
self--energy (Eq.~(\ref{pi1})).
Furthermore, given an additional assumption regarding the relative
contribution of the self--energy insertion to the evaluation of the
full self--energy,
it is seen that the ladder graphs exactly cancel against certain
non--ladder graphs, leaving the one loop term of
Fig.~\ref{oneloop}. This cancellation has been illustrated in some particular
examples in Refs.\cite{smilga,kraemmer,meg}; we have shown that this
feature is, for the most part, algebraic and quite general.

%%%%%%%%%%%%%%%%%%%%%%%%%%%%%%%%%%%%%%%%%%%%%%%%%%%%%%%%%%%%%%%%%%%
\section{Acknowledgments}
We thank E.~Petitgirard and A.~Smilga for valuable discussions.
This work was supported by the Natural Sciences and Engineering Research
Council of Canada.
%%%%%%%%%%%%%%%%%%%%%%%%%%%%%%%%%%%%%%%%%%%%%%%%%%%%%%%%%%%%%%%%%%%
%
% References:
%
%%%%%%%%%%%%%%%%%%%%%%%%%%%%%%%%%%%%%%%%%%%%%%%%%%%%%%%%%%%%%%%%%%%

\begin{figure}
\begin{center}
\leavevmode
\epsfxsize=4 in
\epsfbox{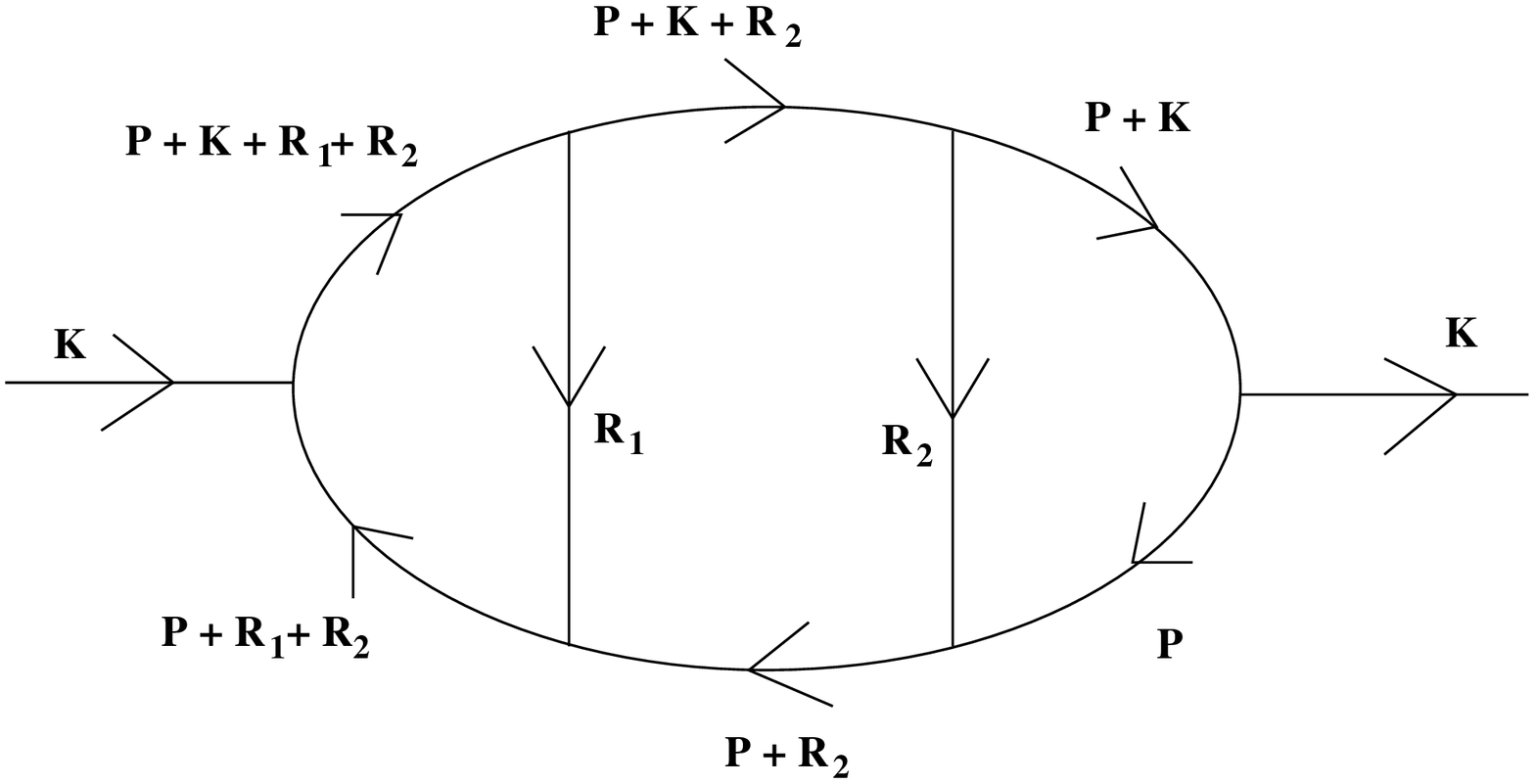}
\end{center}
\caption{A three--loop self--energy ladder graph}
\label{ladderfig}\end{figure}
\begin{figure}
\begin{center}
\leavevmode
\epsfxsize=4 in
\epsfbox{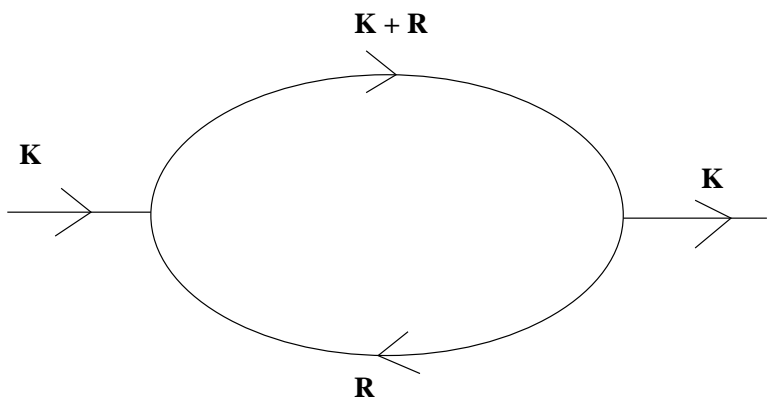}
\end{center}
\caption{A one--loop self--energy graph}
\label{oneloop}\end{figure}
\begin{figure}
\begin{center}
\leavevmode
\epsfxsize=4 in
\epsfbox{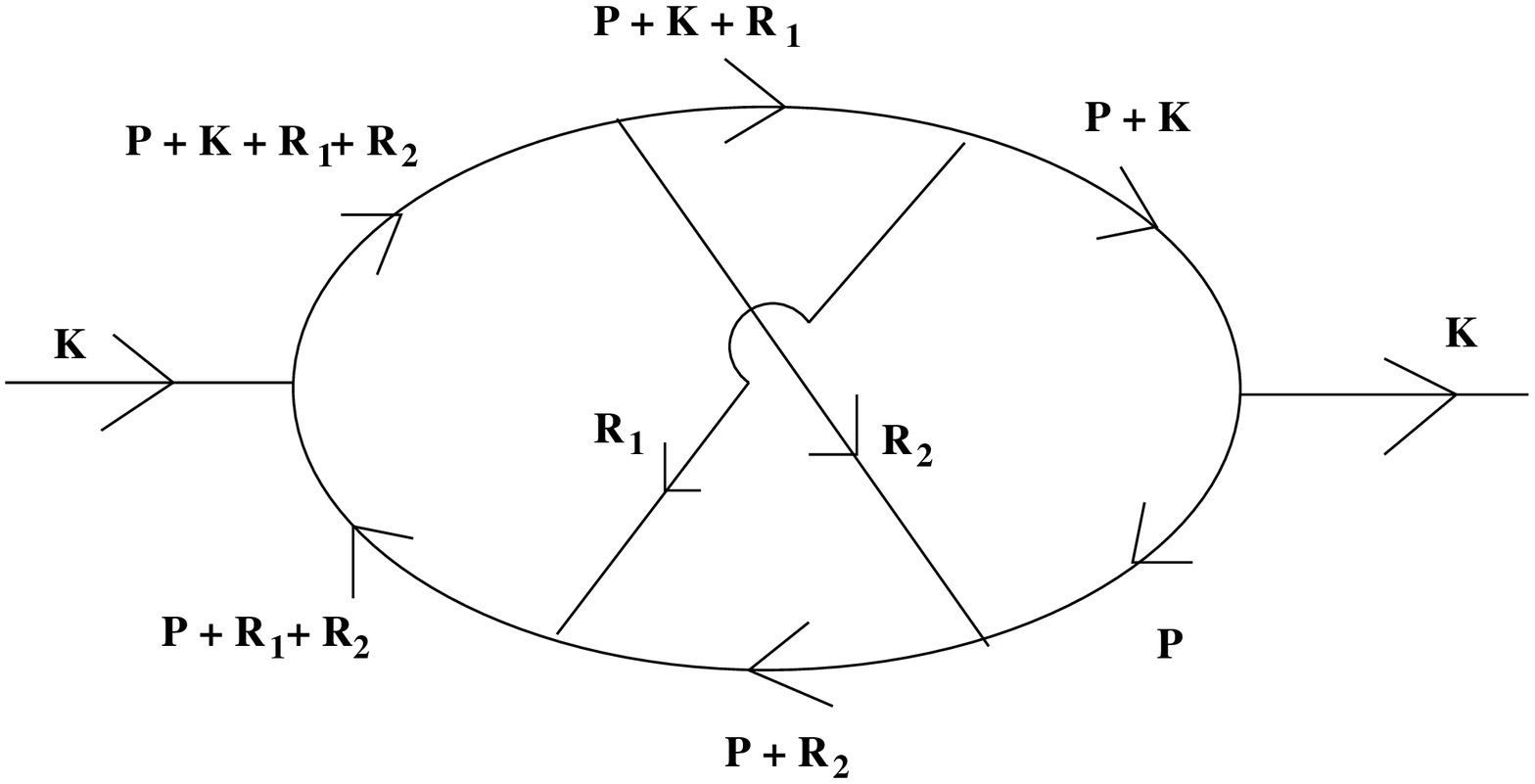}
\end{center}
\caption{A three--loop self--energy non--ladder graph}
\label{unladderfig}\end{figure}
\begin{figure}
\begin{center}
\leavevmode
\epsfxsize=4 in
\epsfbox{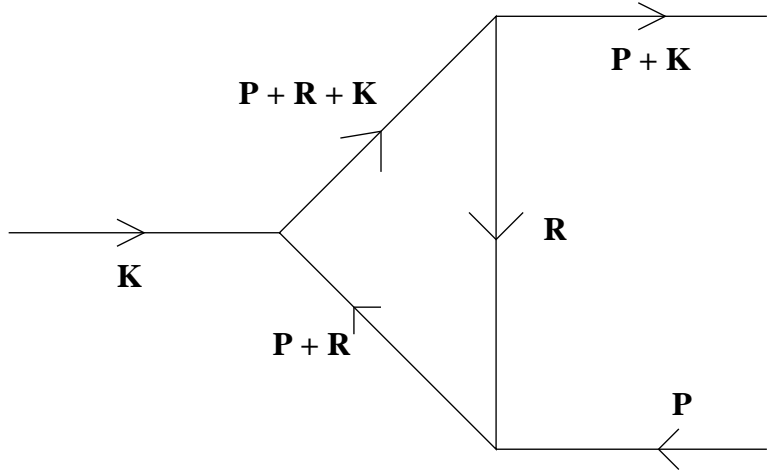}
\end{center}
\caption{A one--loop vertex graph}
\label{vertexfig}\end{figure}
\begin{figure}
\begin{center}
\leavevmode
\epsfxsize=4.5 in
\epsfbox{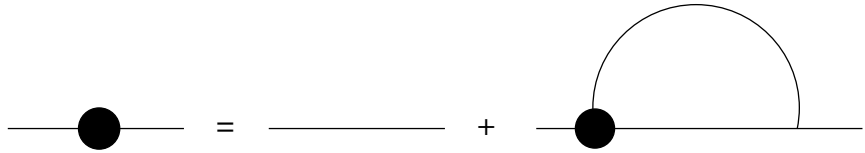}
\end{center}
\caption{A partial Schwinger--Dyson equation for the full self--energy
  which generates the ladder graph resummation}
\label{semiselfenergy}\end{figure}
\begin{figure}
\begin{center}
\leavevmode
\epsfxsize=4.5 in
\epsfbox{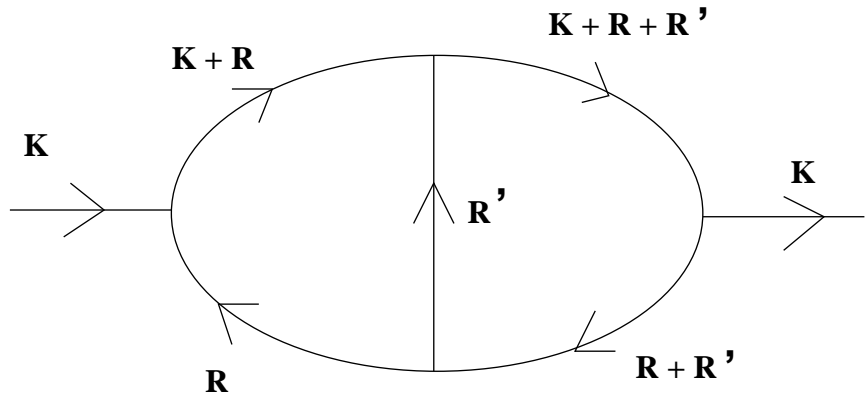}
\end{center}
\caption{A two--loop ladder graph contribution to the self--energy}
\label{firstladder}\end{figure}\clearpage
\begin{figure}
\begin{center}
\leavevmode
\epsfxsize=3.5 in
\epsfbox{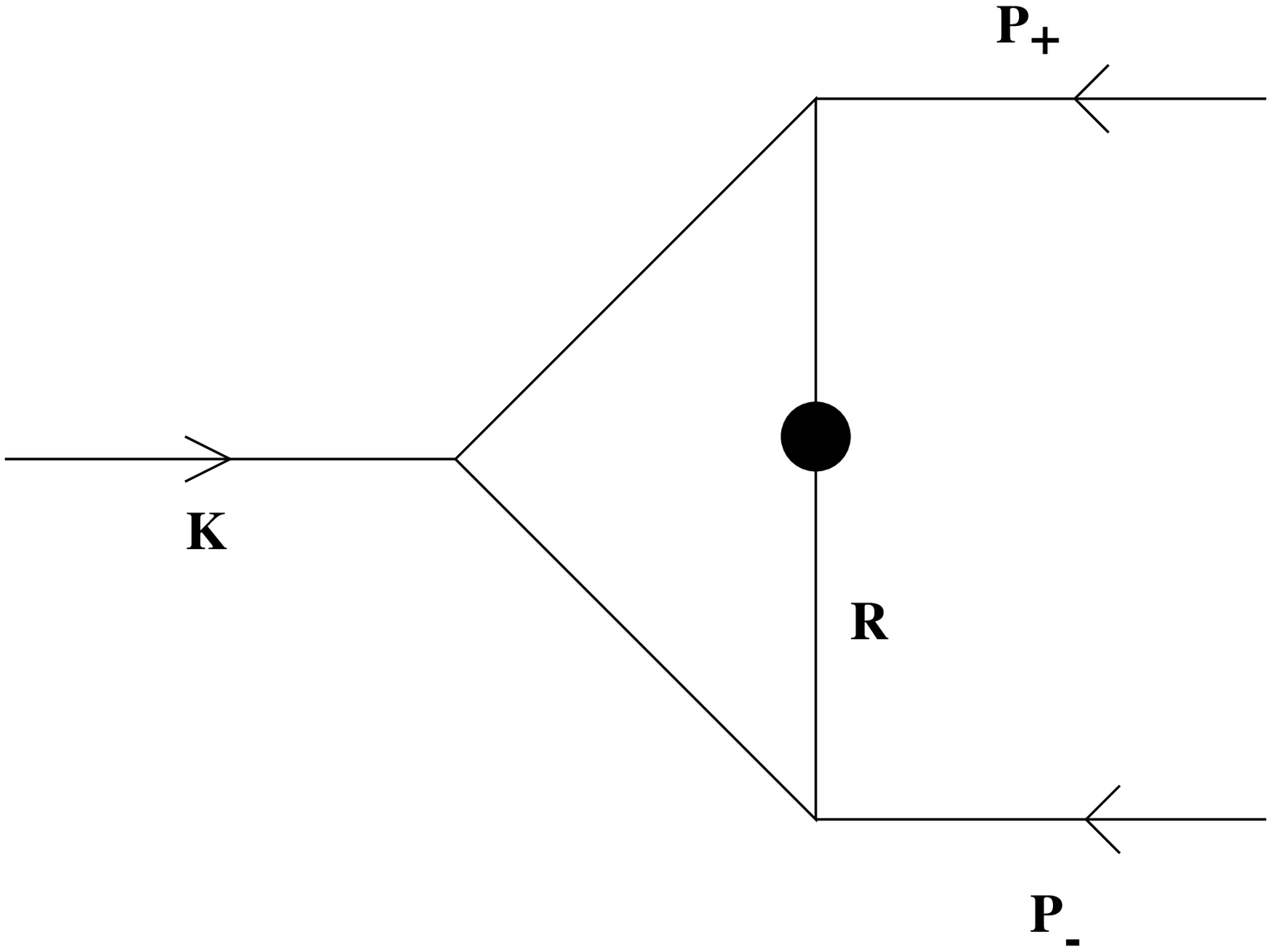}
\end{center}
\caption{Three point vertex with corrected internal line}
\label{fig3}\end{figure}
\begin{figure}
\begin{center}
\leavevmode
\epsfxsize=4 in
\epsfbox{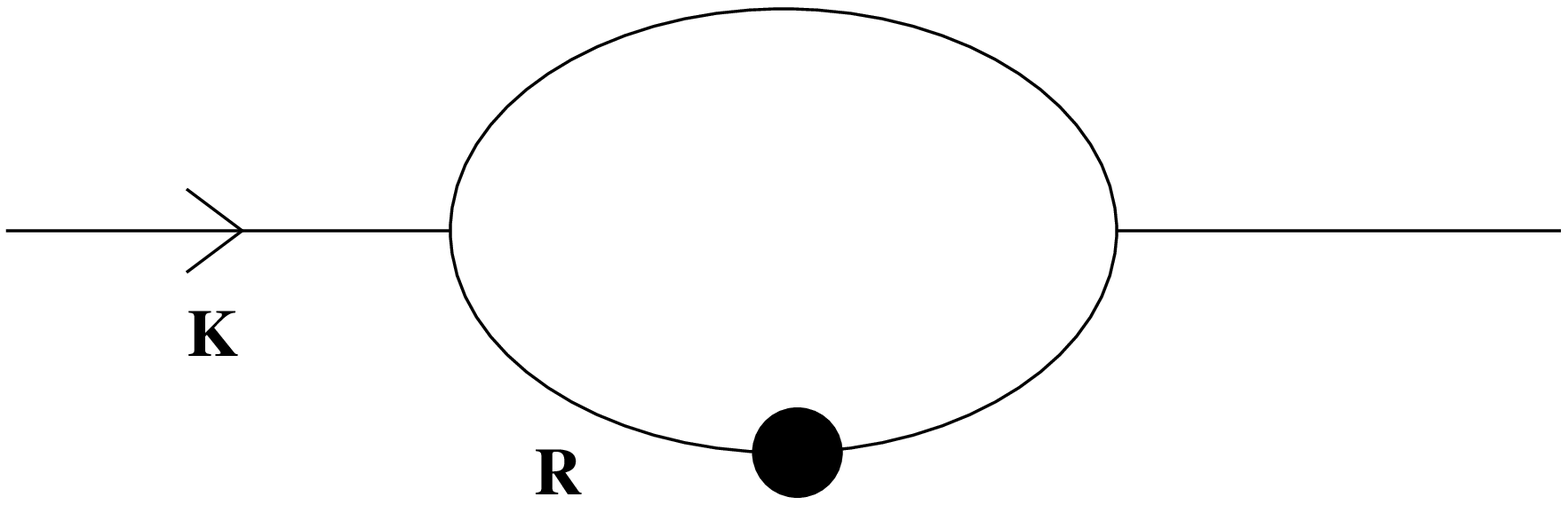}
\end{center}
\caption{Polarization tensor associated with Fig.~\ref{fig3}}
\label{fig4}\end{figure}
\begin{figure}
\begin{center}
\leavevmode
\epsfxsize=4 in
\epsfbox{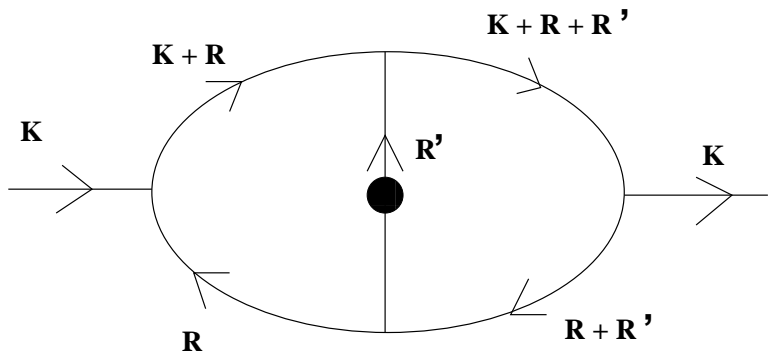}
\end{center}
\caption{Ladder graph corresponding to Fig.~\ref{fig3} with a corrected
internal line}
\label{laddercorrected}\end{figure}\clearpage
\begin{figure}
\begin{center}
\leavevmode
\epsfxsize=4.5 in
\epsfbox{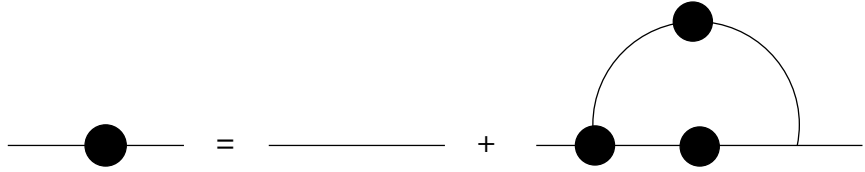}
\end{center}
\caption{The Schwinger--Dyson equation for the full self--energy}
\label{fullselfenergy}\end{figure}
\begin{figure}
\begin{center}
\leavevmode
\epsfxsize=4 in
\epsfbox{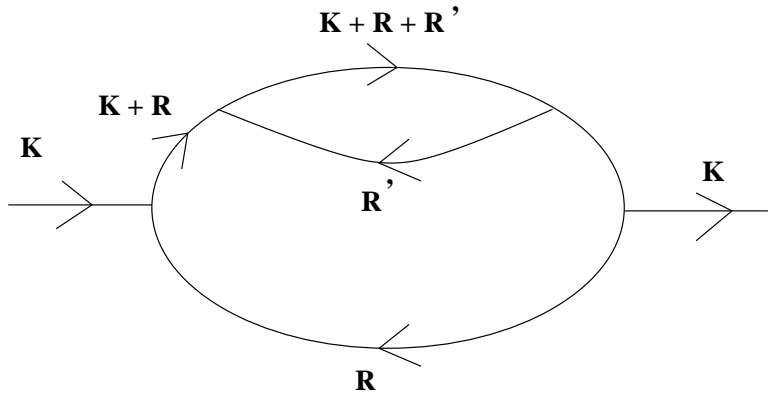}
\end{center}
\caption{Two--loop graph with a self--energy correction}
\label{cancel}\end{figure}
\begin{figure}
\begin{center}
\leavevmode
\epsfxsize=4 in
\epsfbox{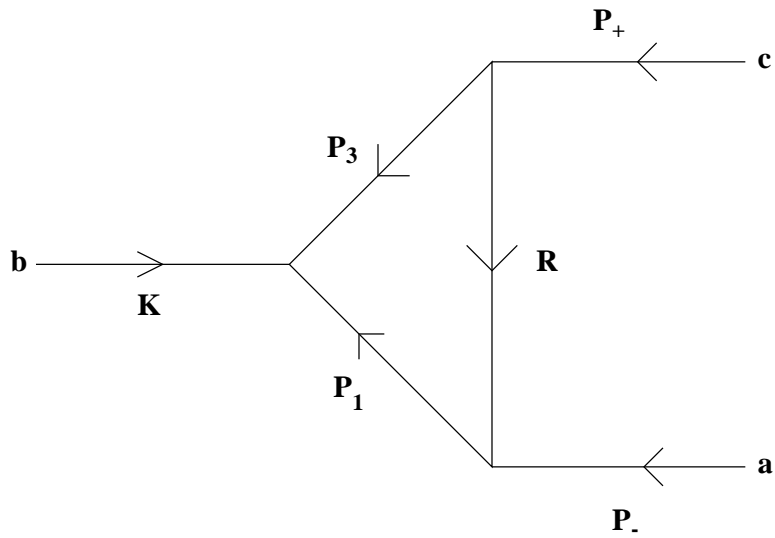}
\end{center}
\caption{Three point vertex}
\label{fig1}\end{figure}
\begin{figure}
\begin{center}
\leavevmode
\epsfxsize=4 in
\epsfbox{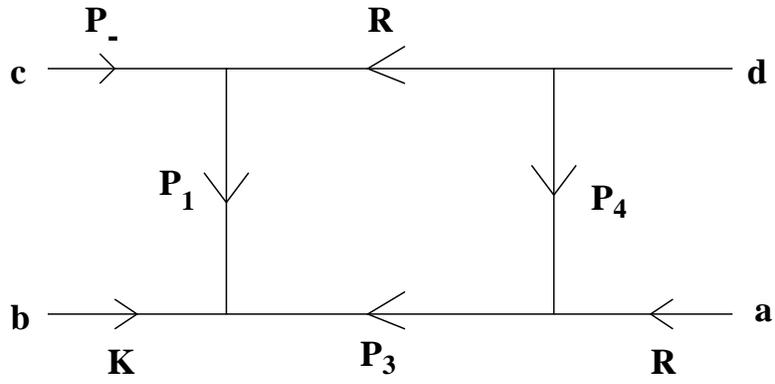}
\end{center}
\caption{Four point vertex}
\label{fig2}\end{figure}
\begin{figure}
\begin{center}
\leavevmode
\epsfxsize=4 in
\epsfbox{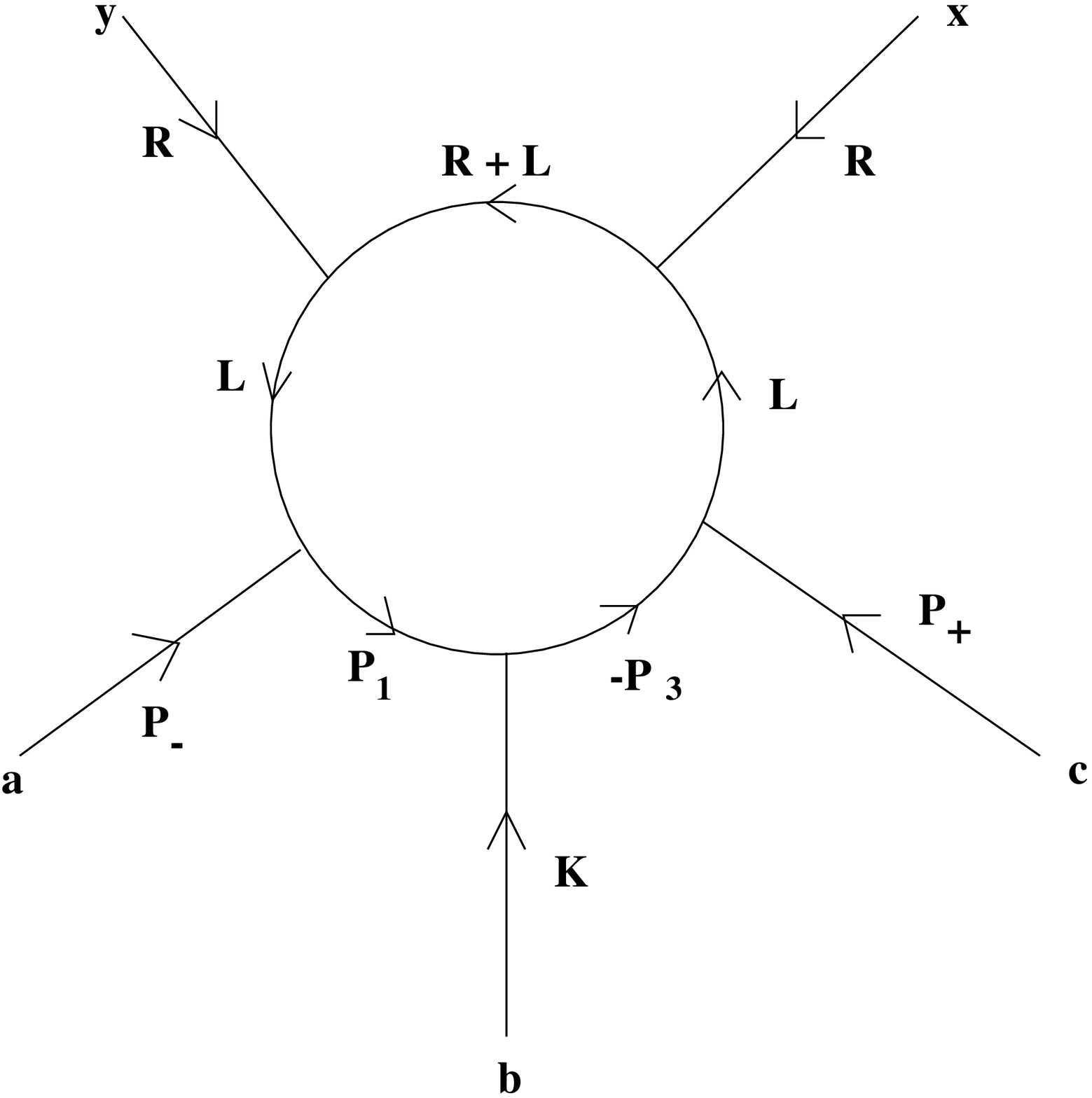}
\end{center}
\caption{Five point vertex graph}
\label{fig2a}\end{figure}\clearpage
\clearpage
\begin{figure}
\begin{center}
\leavevmode
\epsfxsize=2.5 in
\epsfbox{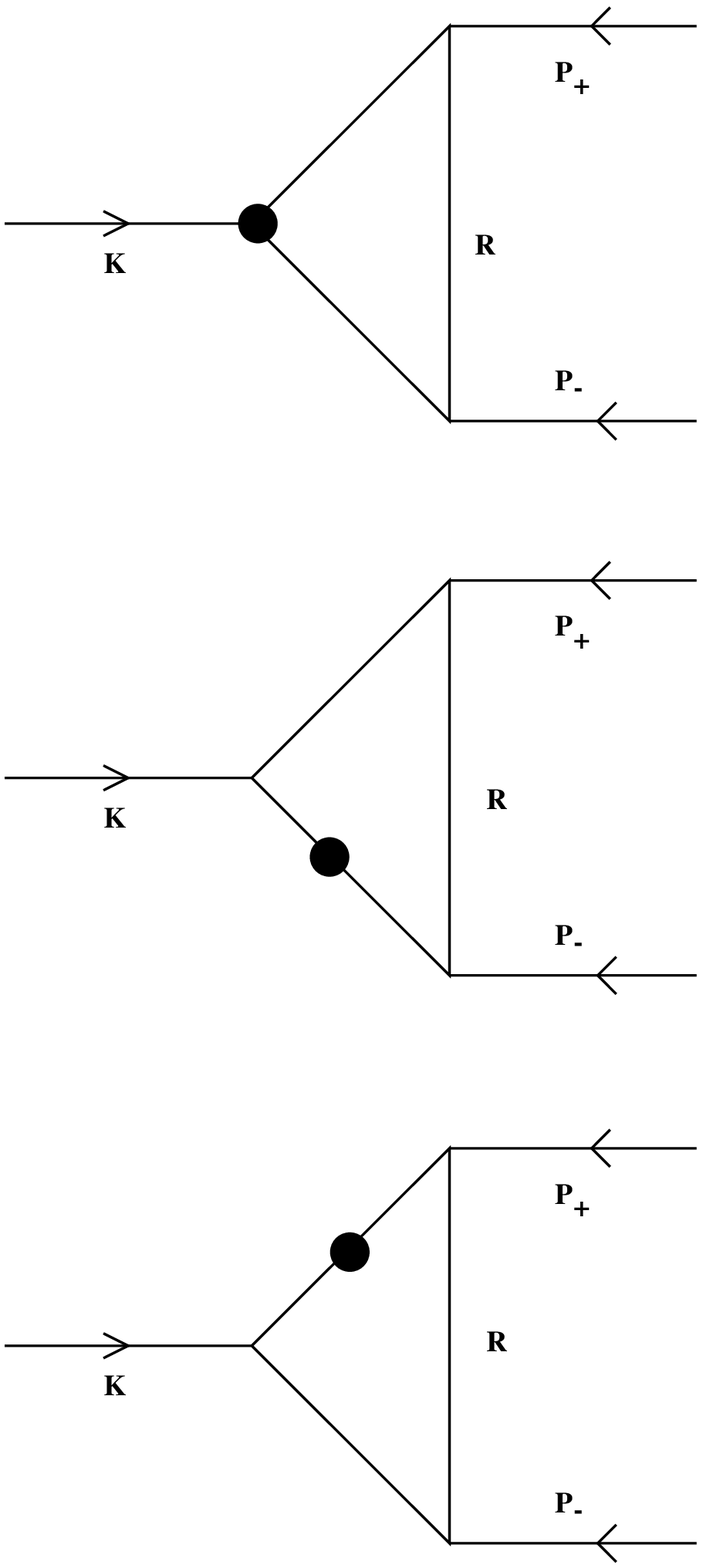}
\end{center}
\caption{Three point vertex with corrected internal line and vertices}
\label{fig5}\end{figure}
\begin{figure}
\begin{center}
\leavevmode
\epsfxsize=4 in
\epsfbox{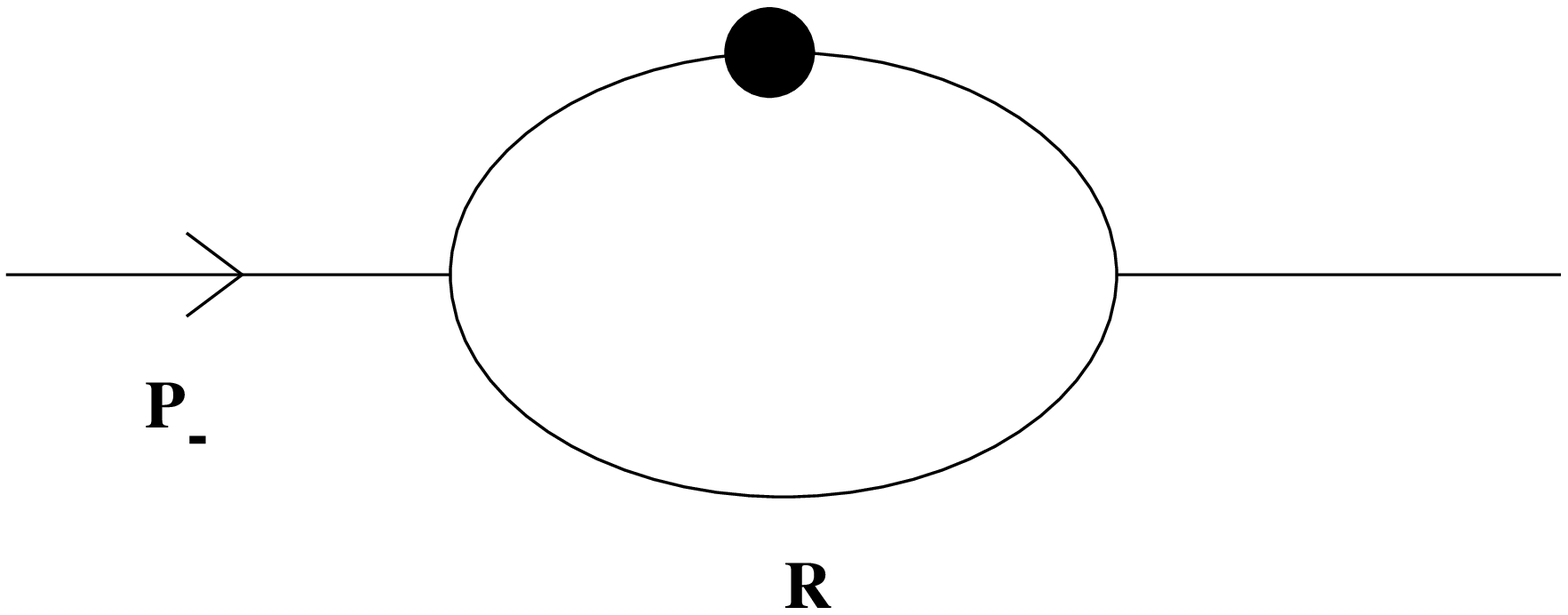}
\end{center}
\caption{Polarization tensor associated with Fig.~\ref{fig5}}
\label{fig6}\end{figure}\clearpage
\begin{figure}
\begin{center}
\leavevmode
\epsfxsize=6 in
\epsfbox{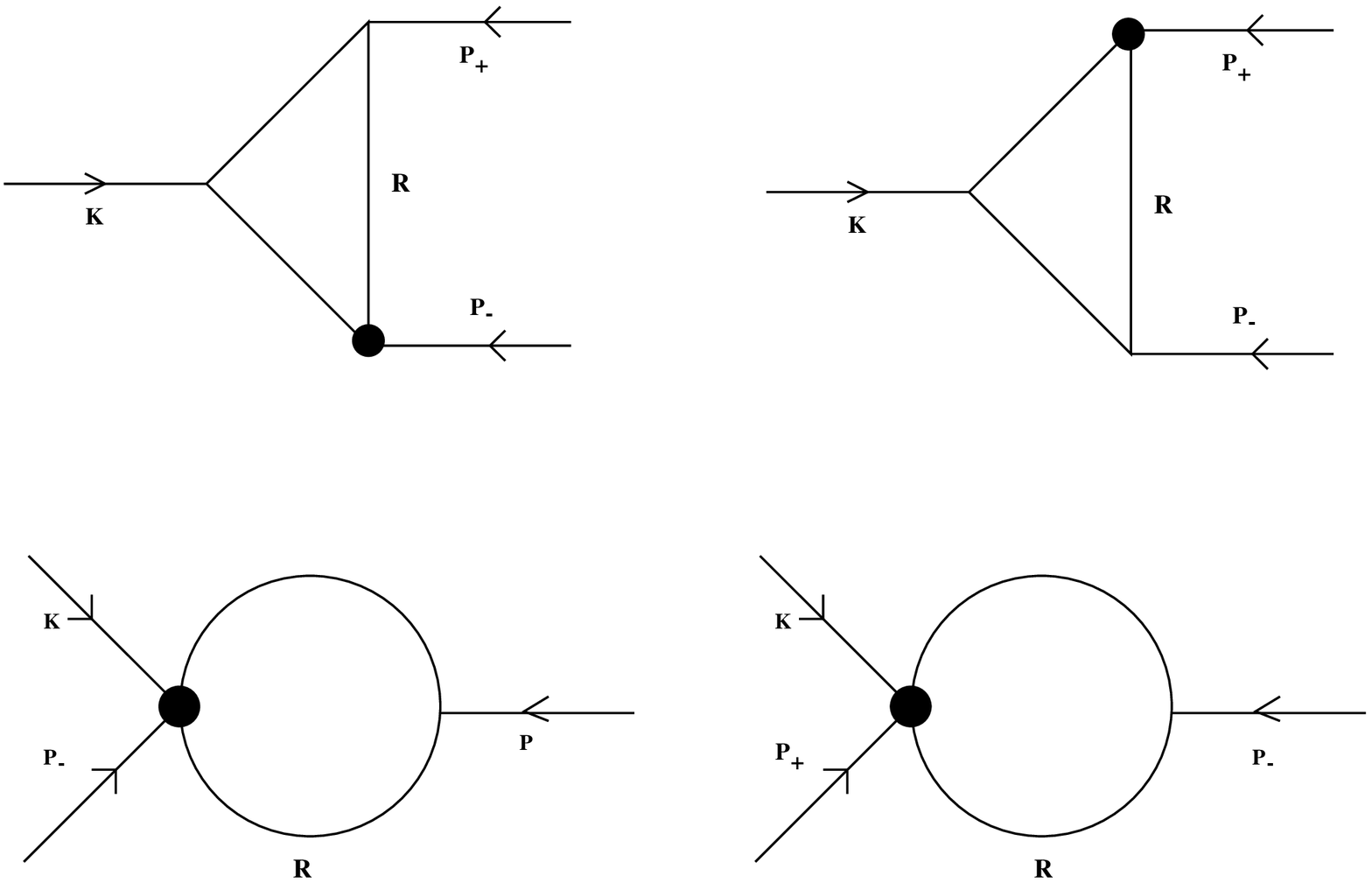}
\end{center}
\caption{Three point vertices with corrected internal vertices}
\label{fig7}\end{figure}
\begin{figure}
\begin{center}
\leavevmode
\epsfxsize=6 in
\epsfbox{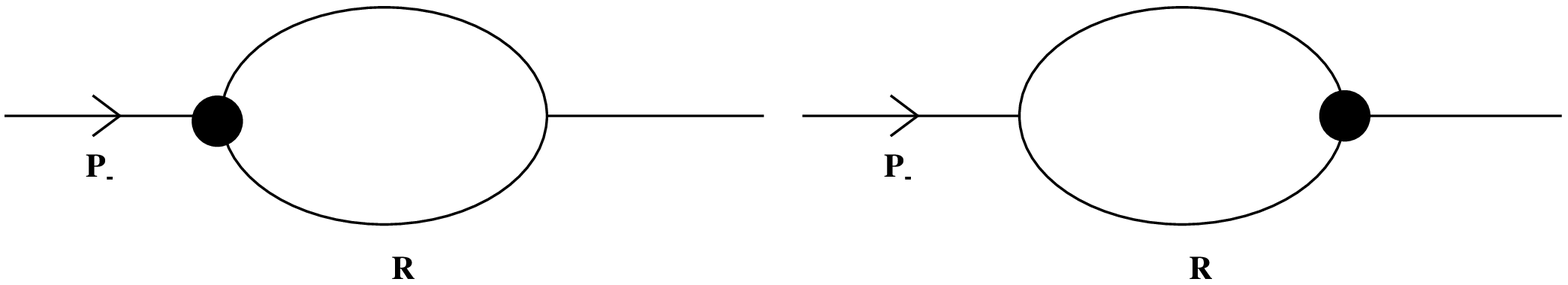}
\end{center}
\caption{Polarization tensor associated with Fig.~\ref{fig7}}
\label{fig8}\end{figure}\clearpage
\begin{figure}
\begin{center}
\leavevmode
\epsfxsize=2.5 in
\epsfbox{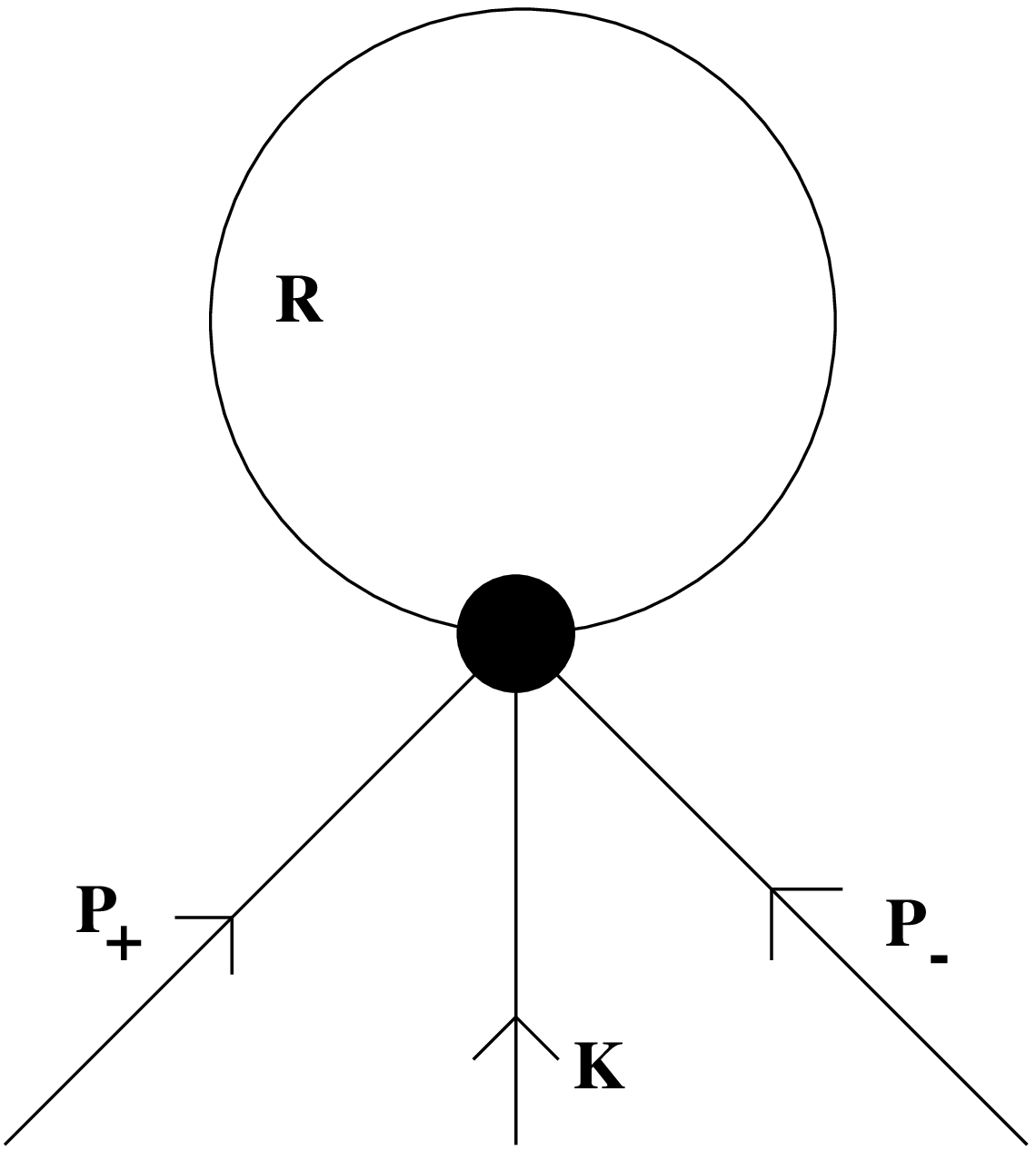}
\end{center}
\caption{Five point vertex}
\label{fig9}\end{figure}
\begin{figure}
\begin{center}
\leavevmode
\epsfxsize=4.2 in
\epsfbox{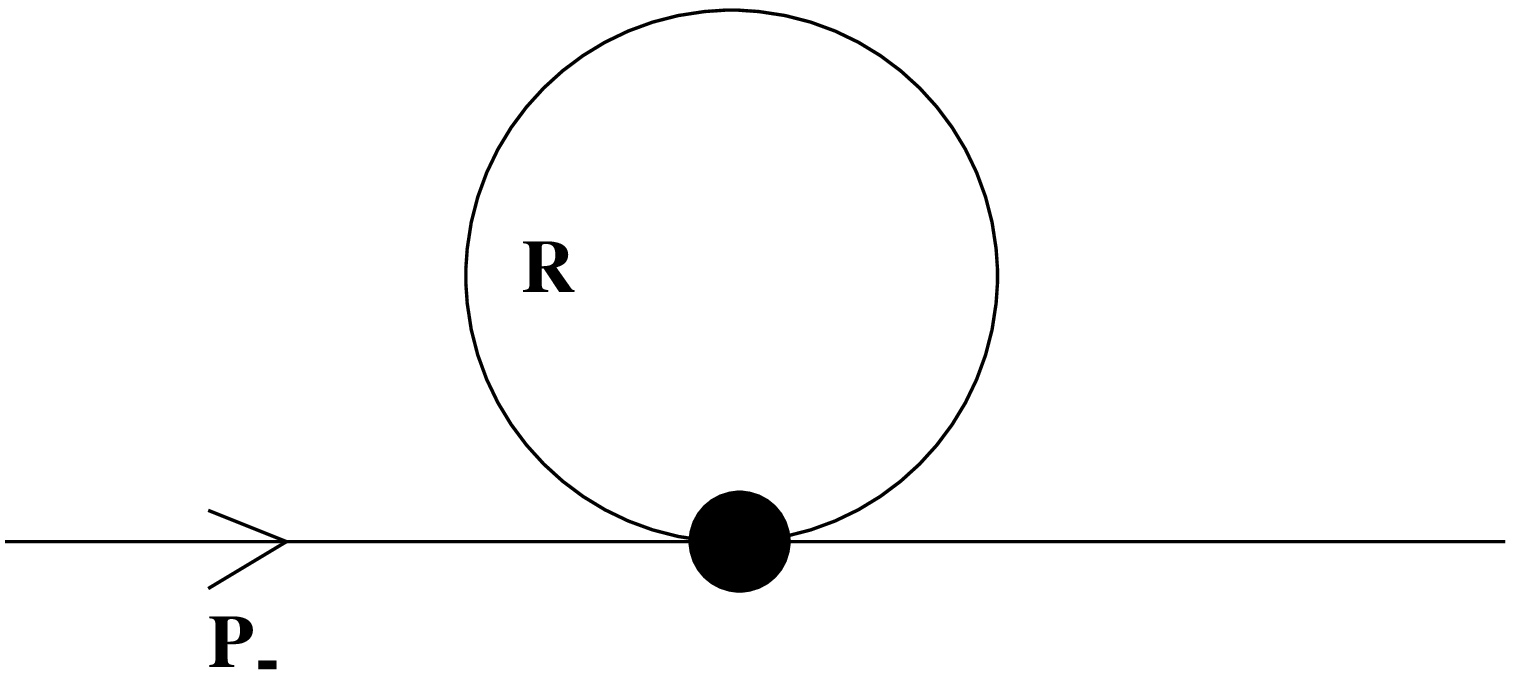}
\end{center}
\caption{Polarization tensor associated with Fig.~\ref{fig9}}
\label{fig10}\end{figure}\clearpage
%\begin{figure}
%\begin{center}
%\leavevmode
%\epsfxsize=4 in
%\epsfbox{./fig11.eps}
%\end{center}
%\caption{Three point vertex with vanishing contribution}
%\label{fig11}\end{figure}
\end{document}